\documentclass[twocolumn,aps,prl,superscriptaddress,amsmath,amssymb]{revtex4-2}

\usepackage{graphicx}
\usepackage{dcolumn}
\usepackage{bm}
\usepackage[T1]{fontenc}
\usepackage{textcomp}

\usepackage{mathptmx}
\usepackage{units}
\usepackage{amsmath}
\usepackage[labelformat=empty]{subfig}
\usepackage{upgreek}
\usepackage{hyperref}
\usepackage{breakurl}
\usepackage{physics}
\usepackage{amsfonts}

\begin{document}
\preprint{AIP/123-QED}
\setlength{\parindent}{0pt}

\title[Parametric Excitation and Instabilities of Spin Waves driven by Surface Acoustic Waves]{Parametric Excitation and Instabilities of Spin Waves driven by Surface Acoustic Waves}

\author{Moritz Geilen}
\affiliation{Fachbereich Physik and Landesforschungszentrum OPTIMAS, Technische Universit\"at Kaiserslautern, Germany}
\email{mgeilen@physik.uni-kl.de}
\author{Roman Verba}
\affiliation{Institute of Magnetism, Kyiv 03142, Ukraine}
\author{Alexandra Nicoloiu}
\affiliation{National Institute for Research and Development in Microtechnologies, Bucharest R-07719, Romania}
\author{Daniele Narducci}
\affiliation{imec, Leuven B-3001, Belgium}
\affiliation{KU Leuven, Departement Materiaalkunde, 3001 Leuven, Belgium}
\author{Adrian Dinescu}
\affiliation{National Institute for Research and Development in Microtechnologies, Bucharest R-07719, Romania}
\author{Milan Ender}
\affiliation{Fachbereich Physik and Landesforschungszentrum OPTIMAS, Technische Universit\"at Kaiserslautern, Germany}
\author{Morteza Mohseni}
\affiliation{Fachbereich Physik and Landesforschungszentrum OPTIMAS, Technische Universit\"at Kaiserslautern, Germany}
\author{Florin Ciubotaru}
\affiliation{imec, Leuven B-3001, Belgium}
\author{Mathias Weiler}
\affiliation{Fachbereich Physik and Landesforschungszentrum OPTIMAS, Technische Universit\"at Kaiserslautern, Germany}
\author{Alexandru M\"uller}
\affiliation{National Institute for Research and Development in Microtechnologies, Bucharest R-07719, Romania}
\author{Burkard Hillebrands}
\affiliation{Fachbereich Physik and Landesforschungszentrum OPTIMAS, Technische Universit\"at Kaiserslautern, Germany}
\author{Christoph Adelmann}
\affiliation{imec, Leuven B-3001, Belgium}
\author{Philipp Pirro}
\affiliation{Fachbereich Physik and Landesforschungszentrum OPTIMAS, Technische Universit\"at Kaiserslautern, Germany}

\date{\today}

\begin{abstract}
The parametric excitation of spin waves by coherent surface acoustic waves is demonstrated experimentally in metallic magnetic thin film structures. The involved magnon modes are analyzed with micro-focused Brillouin light scattering spectroscopy and complementary micromagnetic simulations combined with analytical modelling are used to determine the origin of the spin-wave instabilities. Depending on the experimental conditions, we observe spin-wave instabilities originating from different phonon-magnon and magnon-magnon scattering processes. Our results demonstrate that an efficient excitation of high amplitude, strongly nonlinear magnons in metallic ferromagnets is possible by surface acoustic waves, which opens novel ways to create micro-scaled nonlinear magnonic systems for logic and data processing that can profit from the high excitation efficiency of phonons using piezoelectricity.
\end{abstract}

\maketitle
Magnetoelasticity, the interaction between magnetism and elastic strain, is a promising mechanism for  compact and energy-efficient spintronic devices. In particular, the influence of elastic strain on spin waves, the fundamental excitations in a magnetic solid, is a crucial topic. Many studies have shown that a magnetic field is generated by static strain due to the Villari effect, which can be used to manipulate the ferromagnetic resonance or the propagation of spin waves \cite{Grachev2021,Bukharaev2018,Fetisov2006,Sadovnikov2018}. The interaction between dynamic strain, in the form of bulk acoustic waves (BAWs), and spin waves, and accordingly the formation of hybrid magneto-elastic waves has been studied extensively already in the 1950s and 1960s. As theoretically suggested by Kittel \cite{Kittel1958} the excitation of BAWs by spin waves \cite{Bommel1959} and conversely the excitation of spin waves by BAWs has been demonstrated \cite{Pomerantz1961}. In the linear regime of magneto-elastic coupling, it is proven that the frequency and momentum of the initial wave is transferred during the excitation. In the nonlinear regime, the parametric excitation of BAWs has been shown in high-power ferromagnetic resonance experiments \cite{Schlomann1960}. The parametric interaction of spin waves by BAWs, on the other hand, has been theoretically described \cite{Haas1966} and indirectly demonstrated by Morgenthaler \cite{Matthews1964}. Parametric pumping of magnons by BAWs has been detected by inductive microantennas \cite{Chowdhury2017} and by the inverse spin Hall effect \cite{Polzikova2018,Alekseev2020}.

In addition to BAWs, surface acoustic waves (SAWs) can be used to excite magnetization dynamics by magneto-elastic \cite{Yang2021,Elhosni2016,Thevenard2016} and magneto-rotation coupling \cite{Xu2020}. The symmetry of the magneto-acoustic interaction thereby depends on the SAW mode \cite{Babu2020,Kus2021}. In addition, their spatial localization to the surface makes their interaction with thin (magnetic) films, placed on top of the wave-carrying substrate, much more efficient which opens various new opportunities for hybrid systems and devices \cite{Verba2019SA,Shah2020}.
SAWs are nowadays used in many modern devices such as rf-filters, sensors or lab-on-a-chip applications \cite{Delsing2019,Han2019} due to their efficient and selective excitation on piezoelectric substrates accompanied by very low propagation losses. The excitation of spin waves by SAWs and the associated effects on SAWs have been experimentally demonstrated in various magnetic materials \cite{Weiler2011,Yahagi2017,Kuszewski2018,Ku2020,geilen2021fully}. These works represent a crucial step towards efficient voltage-controlled spin-wave excitation, which is a key requirement for most applications of magnonic circuits. Such magnonic circuits \cite{Mahmoud2020,Pirro2021} implement wave-based logic and are considered a promising complementary technology to CMOS. Moreover, spin waves provide a broad spectrum of nonlinear phenomena \cite{Pirro2014,Wang2020,Schultheiss2012,Krivosik2010}, which opens the door towards novel applications like, e.g., neuromorphic computing \cite{Papp2021,Wang2020}. A crucial feature for neural networks is the presence of non-linear processes with concomitant thresholds. Nonlinear higher harmonic spin-wave generation induced by SAWs has been shown in magnetic semiconductors \cite{Kraimia2020}. However, parametric interaction between spin waves and SAWs as predicted by theory \cite{Lisenkov2019,Zhang2020} has not been experimentally demonstrated.

\begin{figure}[bt]
	\includegraphics[width=8cm]{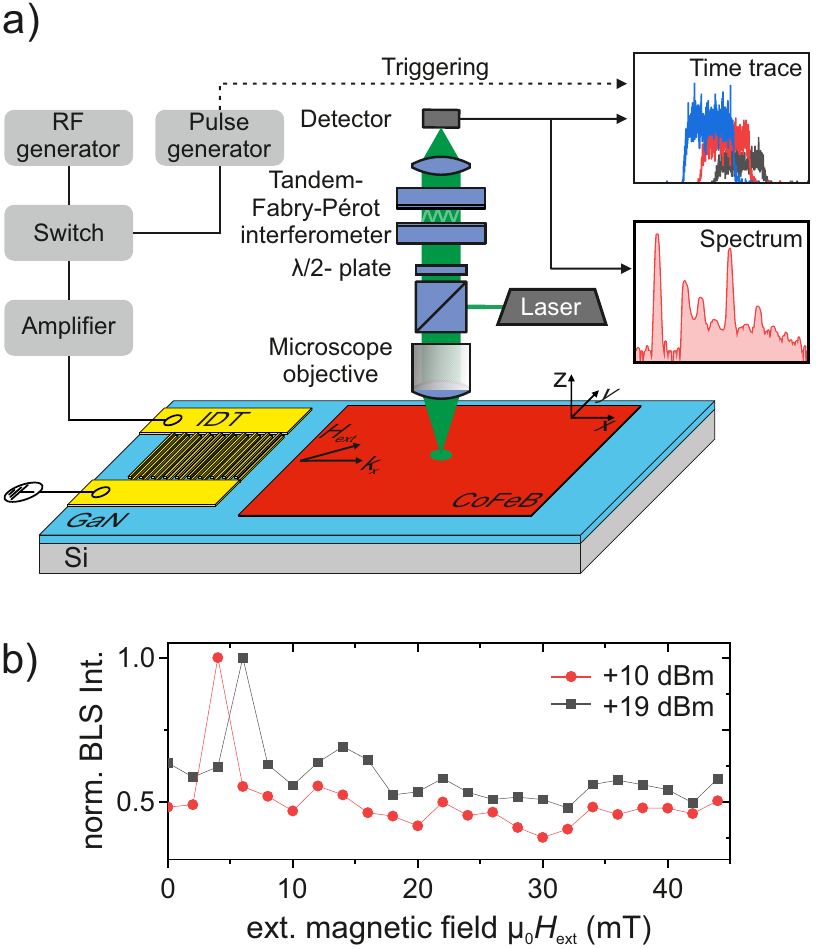}
	\caption{a) Schematic illustration of the experimental setup. b) Normalized BLS intensity at the frequency of the fundamental mode $f_0=\unit[6.3]{GHz}$ as a function of the applied magnetic field for linear and non-linear case. The maxima correspond to the resonant excitation of spin waves by the SAWs.}
	\label{fig:fig1}
\end{figure}

In this letter, we study spin-wave instabilities, in particular the parametric excitation of spin waves by coherent SAWs in a thin metallic ferromagnetic film of CoFeB. Depending on the bias magnetic field, the magnetization dynamics is directly, or alternatively, parametrically driven by the magneto-elastic field generated by SAWs. Micro-focused Brillouin light scattering spectroscopy ($\upmu$BLS, see Fig.~\ref{fig:fig1} a) is used to study the different spin-wave modes in a frequency-resolved manner and to reveal the threshold behaviour of the instability processes. Analytical modelling combined with full micromagnetic simulations is used to reveal the complex excitation schemes of the different instability processes by identifying the involved mechanisms and magnon modes. Depending on the experimental conditions, a four-magnon instability of the magnon mode, linearly driven by the phonon, is observed as well as a direct first-order parametric phonon-to-magnon instability, which is enhanced by three-magnon splitting of the non-resonantly driven magnon, and even three-wave magnon and magnon-phonon confluence processes.

In our experiment, an interdigital transducer (IDT) is used to generate surface acoustic waves with a frequency $f_\mathrm{SAW}=\unit[6.31]{GHz}$ and a wave vector $|k_\mathrm{SAW}|=\unit[9.2]{rad/\upmu m}$ in a $\unit[1.3]{\upmu m}$ thick GaN layer on a silicon substrate. Details on the SAW excitation and propagation can be found in \cite{Geilen2020}. The spin waves are excited in a rectangular pad made from a $\mathrm{Co_{40}Fe_{40}B_{20}}$ thin film with thickness $h_\mathrm{CoFeB}=\unit[18]{nm}$ and are measured by $\upmu$BLS \cite{Sebastian2015,Kargar2021,geilen2021fully}. An external static magnetic bias field is applied in-plane at an angle of $\varphi =\unit[45]{^\circ}$ with respect to the propagation direction of the SAWs (see~Fig.~\ref{fig:fig1}~a). 
For an in-plane magnetized film the excited Rayleigh SAWs create an effective magnetoelastic field $\mu_0 H_\mathrm{mel} = -(2 B_1 S_{xx} M_x/M_s^2 \mathbf{e_x} + B_2 S_{xz} M_x/M_s^2 \mathbf{e_z})$, where $S_{ij}$ is the SAW strain and $B_1$ and $B_2$ are the magneto-elastic coupling constants of CoFeB in units of $\unit[]{MJ/m^3}$.

Recently, we have demonstrated the excitation of spin waves by SAWs in the linear excitation regime \cite{geilen2021fully}. Thereby, a strong increase in the spin-wave intensity is observed at the resonance field, where both the frequency and the wave vector of the SAWs are transferred to the spin wave $(f_\mathrm{SAW},k_\mathrm{SAW})=(f_0,k_0)$. This increase can also be observed in the case presented in this article as shown in Fig.~\ref{fig:fig1}~b). By increasing the RF power, this resonance condition is shifted to higher magnetic fields, due to the nonlinear shift of the spin-wave dispersion \cite{Krivosik2010}.\\

\begin{figure}[bt]
	\includegraphics[width=8cm]{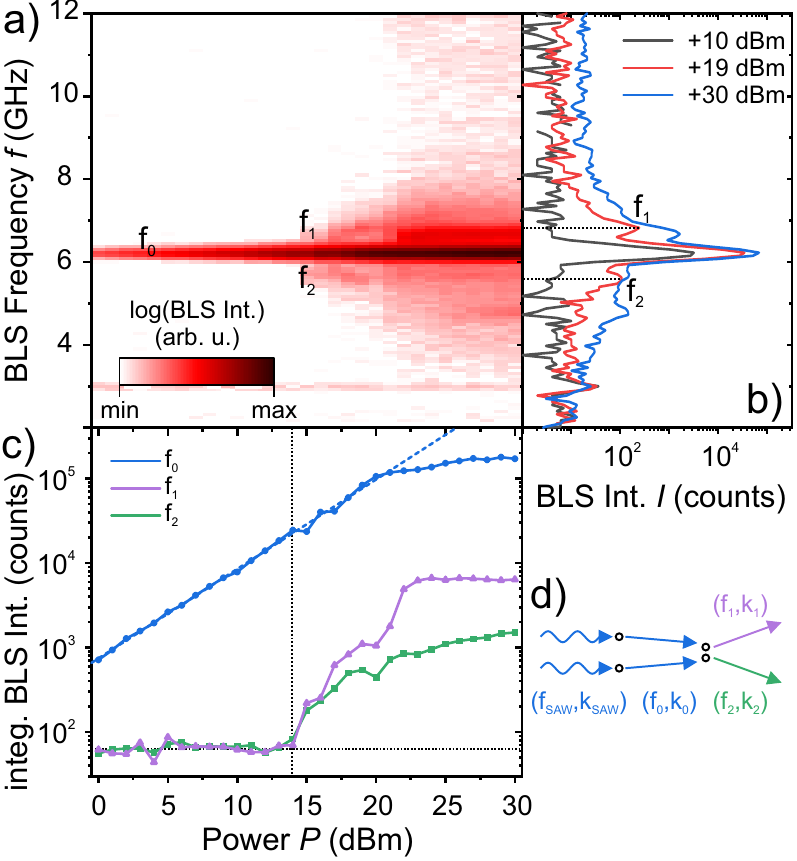}
	\caption{Evolution of BLS spectra with an applied RF power a) and exemplary BLS spectra showing linear, critical and supercritical regimes b) at an external magnetic field of $\upmu_\mathrm{0}H_\mathrm{ext}=\unit[6]{mT}$ on a logarithmic intensity scale. c) Logarithmic intensity for directly excited spin waves at $f_0=\unit[6.3]{GHz}$ (blue) and the intensity of the secondary modes $f_1=\unit[6.9]{GHz}$ (purple) and $f_2=\unit[5.7]{GHz}$ (green). A linear progression is indicated by the blue dashed line and the threshold by the vertical black dotted line. d) Schematic illustration of the second order instability process.}
	\label{fig:fig2}
\end{figure}

First, the case of resonant spin-wave excitation by the magneto-elastic field beyond the linear regime is studied. The magnetic field is kept constant at $\upmu_\mathrm{0}H_\mathrm{ext}=\unit[6]{mT}$ and the BLS spectrum for magnons is measured as a function of the applied RF power (see~Fig.~\ref{fig:fig2}~a). The power $P$ applied to the IDT is varied between $\unit[0]{dBm}$ and $\unit[+30]{dBm}$. For low powers, only the directly excited mode $f_0$ is present. If the power is increased, the appearance of two additional secondary modes $f_1=\unit[5.7]{GHz}$ and $f_2=\unit[6.9]{GHz}$ can be observed close to $f_0$. These two modes satisfy the energy conservation law $2 f_0 = f_1 + f_2$ and, thus, can be attributed to a parametric instability of the second-order \cite{Pirro2014}. For $P\gtrsim\unit[22]{dBm}$ (see~Fig.~\ref{fig:fig2}~b, blue line), spin waves are excited throughout a broad range of the spin-wave band. These spin waves are caused by further, higher-order scattering channels. This region, which is referred to as the supercritical regime \cite{Schultheiss2012,Pirro2014}, will not be considered further. Instead, we will restrict ourselves to the region close to the threshold where only a single instability process occurs.

A characteristic property of nonlinear instability processes is the existence of a threshold in the initial spin-wave mode`s amplitude. This threshold is determined by the minimum of the ratio between the effective relaxation frequency and the coupling strength between the involved magnon modes \cite{Gurevich1996,Stancil2009,Krivosik2010}. From the BLS intensity of the three involved modes, presented in Fig.~\ref{fig:fig2}~c, one can see that the primary mode $f_0$ increases linearly (indicated by dashed line) up to the threshold of the instability process ($P^\mathrm{th}=\unit[+13]{dBm}$). Once this threshold is overcome a sufficient energy flow into the secondary modes appears and the intensity of the initial mode starts to deviate from this linear dependence. Simultaneously, this energy flow leads to an abrupt increase of the intensity of the secondary modes $f_1$ and $f_2$.
It should be mentioned that SAWs themselves can also exhibit nonlinear processes at sufficiently high amplitudes \cite{Mayer2008,Mayer1995}. These would also lead to a deviation from the linear behaviour, which is usually caused by the generation of higher harmonics, which, however, have not been observed in our experiment.

Conclusively, we attribute this observation to the following mechanism: Spin waves are resonantly excited by the SAWs, and subsequently undergo a four-magnon scattering process, leading to the second-order spin-wave instability (see.~Fig.~\ref{fig:fig2}~d). In principle, another four-wave process could be considered, in which two SAWs split directly into two spin waves. This process, however, is very unlikely since the Hamiltonian for such a process is described by $\mathcal{H} \sim a_k^2 c_{k_1}^* c_{k_2}^*$, where $a_k$ and $c_k$ are the SAW and spin-wave amplitudes respectively \cite{Ruckspiegel2014}. As the magneto-elastic energy is a linear function of the strain $S$, which is in turn proportional to $a_k$, this term can be neglected. Such a process may only appear at very high SAW amplitudes, at which the linear approximation of the strain ($S_{ij}={1}/{2}\left({\partial u_i}/{\partial x_j}+{\partial u_j}/{\partial x_i}\right)$) becomes invalid.

\begin{figure}[bt]
	\includegraphics[width=8cm]{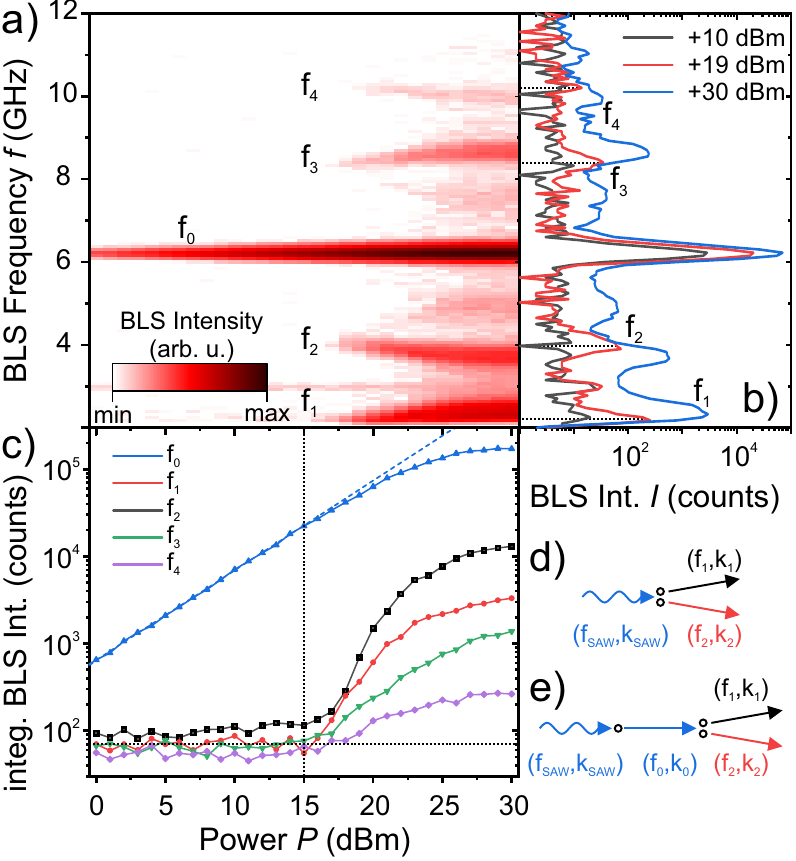}
	\caption{a) Evolution of BLS spectra with an applied RF power a) and exemplary BLS spectra showing linear and critical regimes b) at an external magnetic field of $\upmu_\mathrm{0}H_\mathrm{ext}=\unit[2]{mT}$ on a logarithmic intensity scale. c) Logarithmic intensity at $f_0=\unit[6.3]{GHz}$ (blue) and the intensity of the secondary modes $f_1$ (red), $f_2$ (black), $f_3$ (green) and $f_4$ (purple). A linear progression is indicated by the blue dashed line and the threshold by the vertical black dotted line.Schematic illustration of d) the acoustic pumping process and e) the non-resonant three-magnon process.}
	\label{fig:fig3}
\end{figure}

After considering the case where spin waves are resonantly excited by the SAWs, we will now focus on the non-resonant case. Here, the external magnetic field is reduced to $\upmu_\mathrm{0}H_\mathrm{ext}=\unit[2]{mT}$. As can be seen in Fig.~\ref{fig:fig1}~b, the spin-wave intensity for the directly excited mode at $f_0$ is drastically reduced compared to the resonant case. However, instability processes can be observed in this case as shown in Fig.~\ref{fig:fig3}~a. In the linear regime, only a signal for the initial mode $f_0$ is present, which is non-resonantly driven by the SAWs. At high powers four additional modes with different frequencies $f_1$ to $f_4$ are formed, which are symmetrically spaced around $f_0$. Similar to the case of the second-order instability process, it can be stated that all four secondary modes show a common threshold value ($P^\mathrm{th}=\unit[+15]{dBm}$) within the measurement accuracy. The initial mode starts to deviate from the linear behaviour at this threshold value and begins to saturate (see~Fig.~\ref{fig:fig3}~c).

In the following, the origin of the secondary modes $f_1$ and $f_2$ will be discussed, while the tertiary modes $f_3$ and $f_4$ are adressed in the supplement, since these originate from the interaction between the primary mode and the secondary modes. Near the threshold (see~Fig.~\ref{fig:fig3}~b), the secondary modes have frequencies of $f_1=\unit[2.25]{GHz}$ and $f_2=\unit[4.05]{GHz}$, which satisfy the conservation of energy for a three-wave process $f_0 = f_1 + f_2$ since they are equally spaced around half the frequency of the initial mode $f_0/2$. In materials with reciprocal dispersion relation, this asymmetric splitting is typically observed if the initial mode or correspondingly, the driving field has a non-zero wave vector.

\begin{figure}[bt]
	\includegraphics[width=8cm]{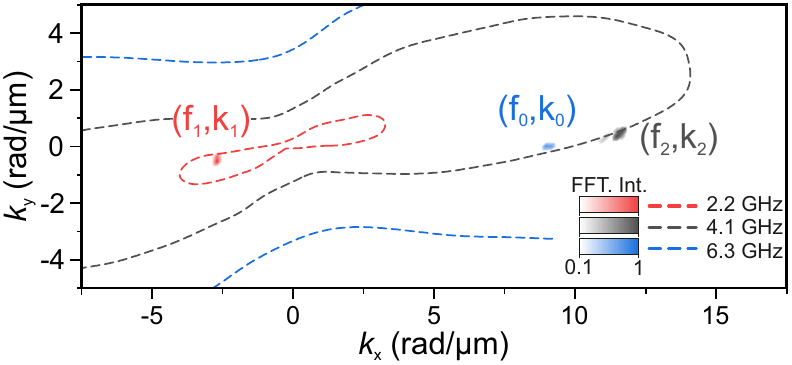}
	\caption{Scheme of first order instability process. The magneto-elastic field is driven with $f_0=\unit[6.3]{GHz}$ (blue) and leads to pumping of the spin-wave modes $f_1=\unit[2.2]{GHz}$ (red) and $f_2=\unit[4.1]{GHz}$ (black) while conserving momentum and energy.}
	\label{fig:sim2}
\end{figure}

In order to determine the wave vectors excited in the observed instability process, micromagnetic simulations were carried out using MuMax3 \cite{Vansteenkiste2014} and the software platform Aithericon \cite{aithericon}. The parameters used for the simulations can be found in the supplemental material \cite{Supplement}. The Rayleigh mode is implemented as a plain wave of the strain components $S_{xx}$ and $S_{xz}$ \cite{Supplement}. 
We find that the magneto-elastic field leads to a forced magnetization dynamics with the frequency and wave vector of the SAW $(f_\mathrm{SAW},k_\mathrm{SAW})=(f_0,k_0)$. As can be seen in Fig.~\ref{fig:sim2} (blue) this is not an eigenmode of the spin wave system since it does not coincide with the isofrequency curve at the SAW frequency (dashed blue line). Once the threshold is exceeded, two secondary spin-wave modes are observed in the simulation: (i) $f_1=\unit[2.2]{GHz} ,~ \mathbf{k}_1= \unit[(-2.5,-0.5)]{rad / \upmu m}$ and (ii) $f_2=\unit[4.1]{GHz} ,~ \mathbf{k}_2= \unit[(11.6,0.5)]{rad / \upmu m}$. These spin-wave modes preserve, within the accuracy of the simulation, the wave vector of the pumping field, in order to fulfill momentum conservation. In addition, there is a good agreement with the experimentally measured frequencies.\\
Based on the findings from the experiment and the micromagnetic simulations, two different three-wave processes are conceivable: (i) \textit{acoustic parametric pumping}, which can be understood in analogy to the parallel pumping of spin waves by electromagnetic waves \cite{Bracher2017} as the scattering of one phonon into two magnons (see~Fig.\ref{fig:fig3}~d), and (ii) \textit{three-magnon splitting of spin waves}, where in our particular case the excitations at $f_0$ are forced excitations by the SAW (see~Fig.\ref{fig:fig3}~e). In order to determine which of the two processes is more efficient, the threshold value was theoretically determined for both cases individually.
The acoustic pumping of spin waves by SAWs is described by the following rate equation and a corresponding equation for $c_2$ \cite{Lisenkov2019}:
\begin{equation}
    \frac{dc_1}{dt} +i\omega_1 c_1 + \Gamma_1 c_1 = i~\mathcal{V}_{\mathrm{SAW},12}~c^*_2~a_\mathrm{SAW}~e^{-i\omega_{\mathrm{SAW}}t},
\end{equation}
where $c_i$ is the spin-wave amplitude of the i-th spin-wave mode, $\Gamma_\mathrm{i}$ the corresponding relaxation rate and $a_\mathrm{SAW}$ the SAW amplitude. The parametric coupling efficiency consists of two contributions and is given by:
\begin{equation}
    \mathcal{V}_{\mathbf{0}, \mathbf{1} \mathbf{2}} = \frac{2\gamma B_1}{M_S}\bar S_{\mathbf{0}, xx} \left(m_{\mathbf{1},x}^* m_{\mathbf{2},x}^* - \mathbf{m}_{\mathbf{1}}^* \cdot \mathbf{m}_{\mathbf{2}}^* \cos^2(\phi_M) \right) \ . 
\end{equation}
Here, $\bar S_{0,xx}$ is the thickness average of the SAW strain $S_{0,xx}$, and $\mathbf{m_k}$ is the spin-wave structure (see details in supplementary \cite{Supplement}). One part is proportional to the ellipticity of the magnetization trajectory, and thus can be interpreted in analogy to the parametric pumping with electromagnetic waves. The other term is characteristic for anisotropy-type pumping, which has been reported for  magneto-elastic \cite{Lisenkov2019} or magneto-electric \cite{Verba2017} drives and depends on the square of the dynamic magnetization components. The minimal threshold for the acoustic pumping $a_\mathrm{th,min}^\mathrm{SAW-SW}$ is found for the splitting into magnons with $\mathbf{k}_1 = \unit[(8.1, -0.5)]{rad/\upmu m}$, $\mathbf{k}_2 = \unit[(1.2, 0.5)]{rad/\upmu m}$ and frequencies of $\unit[4.2]{GHz}$ and $\unit[2.1]{GHz}$. The wavevector of these secondary modes differs significantly from those predicted by the micromagnetic simulations.\\
The second process that needs to be considered is the SAW-driven first order instability, with a forced excitation at $({f_\mathrm{SAW},\mathbf{k}_\mathrm{SAW}})$. 
Like for the acoustic pumping, a rate equation can be formulated to determine the threshold of this instability. For this purpose, the SAW amplitude $a_\mathrm{SAW}$ is replaced by the spin-wave amplitude of the non-resonant mode $c_0$ and the standard three-magnon scattering coefficient $V_{0,12}$ \cite{Krivosik2010} is used as the coupling parameter. The link between $a_\mathrm{SAW}$ and $c_0$ can be found in the Supplement. It turns out that the minimum threshold for this instability process $a_\mathrm{th,min}^\mathrm{3-magnon}$ is about 64\% larger than for the acoustic pumping process. 
Hence, we conclude that the experimentally observed instability is a combination of the two mentioned three-wave processes. The instability is caused by the acoustic pumping of magnons by SAW phonons. However, even below the threshold of the three-magnon instability $a_\mathrm{th,min}^\mathrm{3-magnon}$ an energy flow changes the occupation, and thus, the effective damping of the secondary modes. The combination of these two mechanisms results in the threshold values shown in Fig. 5. The modes with the lowest threshold value for the combined process $a_\mathrm{th,min}^\mathrm{Comb}$ are marked with stars, the ones for pure acoustic pumping with diamonds and the secondary modes obtained by micromagnetic simulations by squares. The small residual discrepancy between the micromagnetic simulations and the combined process can be attributed to the fact that the threshold varies little in this range (less than 6\%). As a result, even small differences in the modeling, i. e., in the calculation of the demagnetization field, can explain this small discrepancy between analytical modeling and micromagnetic simulation.\\
\begin{figure}[bt]
	\includegraphics[width=8cm]{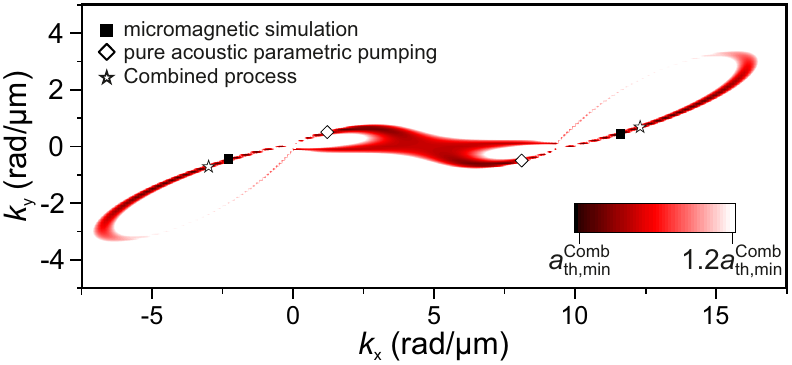}
	\caption{Calculated threshold for the combined process of acoustic pumping process influenced by three-magnon scattering. The spin-wave modes with minimal threshold for the direct SAW-SW process $a_\mathrm{th,min}^\mathrm{SAW-SW}$ (diamond) and the combined process $a_\mathrm{th,min}^\mathrm{Comb}$ (star) are indicated. The modes extracted from the micromagnetic simulations are shown by the black squares.}
	\label{fig:Calculation}
\end{figure}

In summary, in this letter we have studied spin-wave instabilities driven by the magneto-elastic field generated by SAWs. We have investigated two different situations. In the resonant case, where spin waves are excited with the wave vector and the frequency of the SAWs, the second-order instability process due to four-magnon scattering process was observed above a certain SAW amplitude threshold. The second case was the non-resonant excitation of the magnetic system by SAWs. Here, we observed a complex combination of scattering processes, which were identified by a combined study of micromagnetic simulations and analytical modelling. The dominant process is the acoustic parametric pumping process of magnons by the SAWs, which is enhanced by three-magnon scattering channels from the forced excitation at the SAW frequency. This contribution of the three-magnon process is crucial for the determination of the secondary modes. It should be emphasized that in contrast to the case of adiabatic parallel pumping, the initial nonzero wave vector of the SAWs leads to a frequency non-degenerate splitting of the secondary modes. This provides an efficient control of the wave vector spectrum of pumped magnons. Our findings demonstrate that SAW-driven magneto-elastic interaction is sufficiently efficient and large enough to observe various nonlinear magnon and magnon-phonon phenomena, which are prerequisites for magnon information processing systems like phonon-based magnon amplifiers. Thus, our work is a step towards efficient hybrid magnon-phonon systems.\\

\noindent
\begin{acknowledgments}
Financial support by the EU Horizon 2020 research and innovation program within the CHIRON project (contract no. 801055) and the the European Research Council within the Starting Grant No. 101042439 "CoSpiN" is gratefully acknowledged. DN acknowledges financial support from the Research Foundation -Flanders (FWO) through Grants 1SB9121N. R.V. acknowledges support by the Ministry of Education and Science of Ukraine (project No. 0121U110107).
\end{acknowledgments}

%
%
%

\bibliographystyle{apsrev4-2}
\bibliography{Bibtex}

\begin{thebibliography}{51}%
\makeatletter
\providecommand \@ifxundefined [1]{%
 \@ifx{#1\undefined}
}%
\providecommand \@ifnum [1]{%
 \ifnum #1\expandafter \@firstoftwo
 \else \expandafter \@secondoftwo
 \fi
}%
\providecommand \@ifx [1]{%
 \ifx #1\expandafter \@firstoftwo
 \else \expandafter \@secondoftwo
 \fi
}%
\providecommand \natexlab [1]{#1}%
\providecommand \enquote  [1]{``#1''}%
\providecommand \bibnamefont  [1]{#1}%
\providecommand \bibfnamefont [1]{#1}%
\providecommand \citenamefont [1]{#1}%
\providecommand \href@noop [0]{\@secondoftwo}%
\providecommand \href [0]{\begingroup \@sanitize@url \@href}%
\providecommand \@href[1]{\@@startlink{#1}\@@href}%
\providecommand \@@href[1]{\endgroup#1\@@endlink}%
\providecommand \@sanitize@url [0]{\catcode `\\12\catcode `\$12\catcode
  `\&12\catcode `\#12\catcode `\^12\catcode `\_12\catcode `\%12\relax}%
\providecommand \@@startlink[1]{}%
\providecommand \@@endlink[0]{}%
\providecommand \url  [0]{\begingroup\@sanitize@url \@url }%
\providecommand \@url [1]{\endgroup\@href {#1}{\urlprefix }}%
\providecommand \urlprefix  [0]{URL }%
\providecommand \Eprint [0]{\href }%
\providecommand \doibase [0]{https://doi.org/}%
\providecommand \selectlanguage [0]{\@gobble}%
\providecommand \bibinfo  [0]{\@secondoftwo}%
\providecommand \bibfield  [0]{\@secondoftwo}%
\providecommand \translation [1]{[#1]}%
\providecommand \BibitemOpen [0]{}%
\providecommand \bibitemStop [0]{}%
\providecommand \bibitemNoStop [0]{.\EOS\space}%
\providecommand \EOS [0]{\spacefactor3000\relax}%
\providecommand \BibitemShut  [1]{\csname bibitem#1\endcsname}%
\let\auto@bib@innerbib\@empty
\bibitem [{\citenamefont {Grachev}\ \emph {et~al.}(2021)\citenamefont
  {Grachev}, \citenamefont {Matveev}, \citenamefont {Mruczkiewicz},
  \citenamefont {Morozova}, \citenamefont {Beginin}, \citenamefont
  {Sheshukova},\ and\ \citenamefont {Sadovnikov}}]{Grachev2021}%
  \BibitemOpen
  \bibfield  {author} {\bibinfo {author} {\bibfnamefont {A.~A.}\ \bibnamefont
  {Grachev}}, \bibinfo {author} {\bibfnamefont {O.~V.}\ \bibnamefont
  {Matveev}}, \bibinfo {author} {\bibfnamefont {M.}~\bibnamefont
  {Mruczkiewicz}}, \bibinfo {author} {\bibfnamefont {M.~A.}\ \bibnamefont
  {Morozova}}, \bibinfo {author} {\bibfnamefont {E.~N.}\ \bibnamefont
  {Beginin}}, \bibinfo {author} {\bibfnamefont {S.~E.}\ \bibnamefont
  {Sheshukova}},\ and\ \bibinfo {author} {\bibfnamefont {A.~V.}\ \bibnamefont
  {Sadovnikov}},\ }\href {https://doi.org/10.1063/5.0051429} {\bibfield
  {journal} {\bibinfo  {journal} {Applied Physics Letters}\ }\textbf {\bibinfo
  {volume} {118}},\ \bibinfo {pages} {262405} (\bibinfo {year}
  {2021})}\BibitemShut {NoStop}%
\bibitem [{\citenamefont {Bukharaev}\ \emph {et~al.}(2018)\citenamefont
  {Bukharaev}, \citenamefont {Zvezdin}, \citenamefont {Pyatakov},\ and\
  \citenamefont {Fetisov}}]{Bukharaev2018}%
  \BibitemOpen
  \bibfield  {author} {\bibinfo {author} {\bibfnamefont {A.}~\bibnamefont
  {Bukharaev}}, \bibinfo {author} {\bibfnamefont {A.~K.}\ \bibnamefont
  {Zvezdin}}, \bibinfo {author} {\bibfnamefont {A.~P.}\ \bibnamefont
  {Pyatakov}},\ and\ \bibinfo {author} {\bibfnamefont {Y.~K.}\ \bibnamefont
  {Fetisov}},\ }\href {https://doi.org/10.3367/ufnr.2018.01.038279} {\bibfield
  {journal} {\bibinfo  {journal} {Phys. Usp.}\ }\textbf {\bibinfo {volume}
  {61}},\ \bibinfo {pages} {1175} (\bibinfo {year} {2018})}\BibitemShut
  {NoStop}%
\bibitem [{\citenamefont {Fetisov}\ and\ \citenamefont
  {Srinivasan}(2006)}]{Fetisov2006}%
  \BibitemOpen
  \bibfield  {author} {\bibinfo {author} {\bibfnamefont {Y.~K.}\ \bibnamefont
  {Fetisov}}\ and\ \bibinfo {author} {\bibfnamefont {G.}~\bibnamefont
  {Srinivasan}},\ }\href {https://doi.org/10.1063/1.2191950} {\bibfield
  {journal} {\bibinfo  {journal} {Applied Physics Letters}\ }\textbf {\bibinfo
  {volume} {88}},\ \bibinfo {pages} {143503} (\bibinfo {year}
  {2006})}\BibitemShut {NoStop}%
\bibitem [{\citenamefont {Sadovnikov}\ \emph {et~al.}(2018)\citenamefont
  {Sadovnikov}, \citenamefont {Grachev}, \citenamefont {Sheshukova},
  \citenamefont {Sharaevskii}, \citenamefont {Serdobintsev}, \citenamefont
  {Mitin},\ and\ \citenamefont {Nikitov}}]{Sadovnikov2018}%
  \BibitemOpen
  \bibfield  {author} {\bibinfo {author} {\bibfnamefont {A.~V.}\ \bibnamefont
  {Sadovnikov}}, \bibinfo {author} {\bibfnamefont {A.~A.}\ \bibnamefont
  {Grachev}}, \bibinfo {author} {\bibfnamefont {S.~E.}\ \bibnamefont
  {Sheshukova}}, \bibinfo {author} {\bibfnamefont {Y.~P.}\ \bibnamefont
  {Sharaevskii}}, \bibinfo {author} {\bibfnamefont {A.~A.}\ \bibnamefont
  {Serdobintsev}}, \bibinfo {author} {\bibfnamefont {D.~M.}\ \bibnamefont
  {Mitin}},\ and\ \bibinfo {author} {\bibfnamefont {S.~A.}\ \bibnamefont
  {Nikitov}},\ }\href {https://doi.org/10.1103/PhysRevLett.120.257203}
  {\bibfield  {journal} {\bibinfo  {journal} {Physical Review Letters}\
  }\textbf {\bibinfo {volume} {120}},\ \bibinfo {pages} {257203} (\bibinfo
  {year} {2018})}\BibitemShut {NoStop}%
\bibitem [{\citenamefont {Kittel}(1958)}]{Kittel1958}%
  \BibitemOpen
  \bibfield  {author} {\bibinfo {author} {\bibfnamefont {C.}~\bibnamefont
  {Kittel}},\ }\href {https://doi.org/10.1103/PhysRev.110.836} {\bibfield
  {journal} {\bibinfo  {journal} {Physical Review}\ }\textbf {\bibinfo {volume}
  {110}},\ \bibinfo {pages} {836} (\bibinfo {year} {1958})}\BibitemShut
  {NoStop}%
\bibitem [{\citenamefont {B{\"{o}}mmel}\ and\ \citenamefont
  {Dransfeld}(1959)}]{Bommel1959}%
  \BibitemOpen
  \bibfield  {author} {\bibinfo {author} {\bibfnamefont {H.}~\bibnamefont
  {B{\"{o}}mmel}}\ and\ \bibinfo {author} {\bibfnamefont {K.}~\bibnamefont
  {Dransfeld}},\ }\href {https://doi.org/10.1103/PhysRevLett.3.83} {\bibfield
  {journal} {\bibinfo  {journal} {Physical Review Letters}\ }\textbf {\bibinfo
  {volume} {3}},\ \bibinfo {pages} {83} (\bibinfo {year} {1959})}\BibitemShut
  {NoStop}%
\bibitem [{\citenamefont {Pomerantz}(1961)}]{Pomerantz1961}%
  \BibitemOpen
  \bibfield  {author} {\bibinfo {author} {\bibfnamefont {M.}~\bibnamefont
  {Pomerantz}},\ }\href {https://doi.org/10.1103/PhysRevLett.7.312} {\bibfield
  {journal} {\bibinfo  {journal} {Physical Review Letters}\ }\textbf {\bibinfo
  {volume} {7}},\ \bibinfo {pages} {312} (\bibinfo {year} {1961})}\BibitemShut
  {NoStop}%
\bibitem [{\citenamefont {Schl{\"{o}}mann}(1960)}]{Schlomann1960}%
  \BibitemOpen
  \bibfield  {author} {\bibinfo {author} {\bibfnamefont {E.}~\bibnamefont
  {Schl{\"{o}}mann}},\ }\href {https://doi.org/10.1063/1.1735909} {\bibfield
  {journal} {\bibinfo  {journal} {Journal of Applied Physics}\ }\textbf
  {\bibinfo {volume} {31}},\ \bibinfo {pages} {1647} (\bibinfo {year}
  {1960})}\BibitemShut {NoStop}%
\bibitem [{\citenamefont {Haas}(1966)}]{Haas1966}%
  \BibitemOpen
  \bibfield  {author} {\bibinfo {author} {\bibfnamefont {C.~W.}\ \bibnamefont
  {Haas}},\ }\href@noop {} {\bibfield  {journal} {\bibinfo  {journal} {Journal
  of Physics and Chemistry of Solids}\ }\textbf {\bibinfo {volume} {27}},\
  \bibinfo {pages} {1687} (\bibinfo {year} {1966})}\BibitemShut {NoStop}%
\bibitem [{\citenamefont {Matthews}\ and\ \citenamefont
  {Morgenthaler}(1964)}]{Matthews1964}%
  \BibitemOpen
  \bibfield  {author} {\bibinfo {author} {\bibfnamefont {H.}~\bibnamefont
  {Matthews}}\ and\ \bibinfo {author} {\bibfnamefont {F.~R.}\ \bibnamefont
  {Morgenthaler}},\ }\href {https://doi.org/10.1103/PhysRevLett.13.614}
  {\bibfield  {journal} {\bibinfo  {journal} {Physical Review Letters}\
  }\textbf {\bibinfo {volume} {13}},\ \bibinfo {pages} {614} (\bibinfo {year}
  {1964})}\BibitemShut {NoStop}%
\bibitem [{\citenamefont {Chowdhury}\ \emph {et~al.}(2017)\citenamefont
  {Chowdhury}, \citenamefont {Jander},\ and\ \citenamefont
  {Dhagat}}]{Chowdhury2017}%
  \BibitemOpen
  \bibfield  {author} {\bibinfo {author} {\bibfnamefont {P.}~\bibnamefont
  {Chowdhury}}, \bibinfo {author} {\bibfnamefont {A.}~\bibnamefont {Jander}},\
  and\ \bibinfo {author} {\bibfnamefont {P.}~\bibnamefont {Dhagat}},\ }\href
  {https://doi.org/10.1109/LMAG.2017.2737962} {\bibfield  {journal} {\bibinfo
  {journal} {IEEE Magnetics Letters}\ }\textbf {\bibinfo {volume} {8}},\
  \bibinfo {pages} {1} (\bibinfo {year} {2017})}\BibitemShut {NoStop}%
\bibitem [{\citenamefont {Polzikova}\ \emph {et~al.}(2018)\citenamefont
  {Polzikova}, \citenamefont {Alekseev}, \citenamefont {Pyataikin},
  \citenamefont {Luzanov}, \citenamefont {Raevskiy},\ and\ \citenamefont
  {Kotov}}]{Polzikova2018}%
  \BibitemOpen
  \bibfield  {author} {\bibinfo {author} {\bibfnamefont {N.~I.}\ \bibnamefont
  {Polzikova}}, \bibinfo {author} {\bibfnamefont {S.~G.}\ \bibnamefont
  {Alekseev}}, \bibinfo {author} {\bibfnamefont {I.~I.}\ \bibnamefont
  {Pyataikin}}, \bibinfo {author} {\bibfnamefont {V.~A.}\ \bibnamefont
  {Luzanov}}, \bibinfo {author} {\bibfnamefont {A.~O.}\ \bibnamefont
  {Raevskiy}},\ and\ \bibinfo {author} {\bibfnamefont {V.~A.}\ \bibnamefont
  {Kotov}},\ }\href {https://doi.org/10.1063/1.5007685} {\bibfield  {journal}
  {\bibinfo  {journal} {AIP Advances}\ }\textbf {\bibinfo {volume} {8}},\
  \bibinfo {pages} {056128} (\bibinfo {year} {2018})}\BibitemShut {NoStop}%
\bibitem [{\citenamefont {Alekseev}\ \emph {et~al.}(2020)\citenamefont
  {Alekseev}, \citenamefont {Dizhur}, \citenamefont {Polzikova}, \citenamefont
  {Luzanov}, \citenamefont {Raevskiy}, \citenamefont {Orlov}, \citenamefont
  {Kotov},\ and\ \citenamefont {Nikitov}}]{Alekseev2020}%
  \BibitemOpen
  \bibfield  {author} {\bibinfo {author} {\bibfnamefont {S.~G.}\ \bibnamefont
  {Alekseev}}, \bibinfo {author} {\bibfnamefont {S.~E.}\ \bibnamefont
  {Dizhur}}, \bibinfo {author} {\bibfnamefont {N.~I.}\ \bibnamefont
  {Polzikova}}, \bibinfo {author} {\bibfnamefont {V.~A.}\ \bibnamefont
  {Luzanov}}, \bibinfo {author} {\bibfnamefont {A.~O.}\ \bibnamefont
  {Raevskiy}}, \bibinfo {author} {\bibfnamefont {A.~P.}\ \bibnamefont {Orlov}},
  \bibinfo {author} {\bibfnamefont {V.~A.}\ \bibnamefont {Kotov}},\ and\
  \bibinfo {author} {\bibfnamefont {S.~A.}\ \bibnamefont {Nikitov}},\ }\href
  {https://doi.org/10.1063/5.0022267} {\bibfield  {journal} {\bibinfo
  {journal} {Applied Physics Letters}\ }\textbf {\bibinfo {volume} {117}},\
  \bibinfo {pages} {072408} (\bibinfo {year} {2020})}\BibitemShut {NoStop}%
\bibitem [{\citenamefont {Yang}\ and\ \citenamefont
  {Schmidt}(2021)}]{Yang2021}%
  \BibitemOpen
  \bibfield  {author} {\bibinfo {author} {\bibfnamefont {W.~G.}\ \bibnamefont
  {Yang}}\ and\ \bibinfo {author} {\bibfnamefont {H.}~\bibnamefont {Schmidt}},\
  }\href {https://doi.org/10.1063/5.0042138} {\bibfield  {journal} {\bibinfo
  {journal} {Applied Physics Reviews}\ }\textbf {\bibinfo {volume} {8}},\
  \bibinfo {pages} {021304} (\bibinfo {year} {2021})}\BibitemShut {NoStop}%
\bibitem [{\citenamefont {Elhosni}\ \emph {et~al.}(2016)\citenamefont
  {Elhosni}, \citenamefont {Elmazria}, \citenamefont {Petit-Watelot},
  \citenamefont {Bouvot}, \citenamefont {Zhgoon}, \citenamefont {Talbi},
  \citenamefont {Hehn}, \citenamefont {Aissa}, \citenamefont {Hage-Ali},
  \citenamefont {Lacour}, \citenamefont {Sarry},\ and\ \citenamefont
  {Boumatar}}]{Elhosni2016}%
  \BibitemOpen
  \bibfield  {author} {\bibinfo {author} {\bibfnamefont {M.}~\bibnamefont
  {Elhosni}}, \bibinfo {author} {\bibfnamefont {O.}~\bibnamefont {Elmazria}},
  \bibinfo {author} {\bibfnamefont {S.}~\bibnamefont {Petit-Watelot}}, \bibinfo
  {author} {\bibfnamefont {L.}~\bibnamefont {Bouvot}}, \bibinfo {author}
  {\bibfnamefont {S.}~\bibnamefont {Zhgoon}}, \bibinfo {author} {\bibfnamefont
  {A.}~\bibnamefont {Talbi}}, \bibinfo {author} {\bibfnamefont
  {M.}~\bibnamefont {Hehn}}, \bibinfo {author} {\bibfnamefont {K.~A.}\
  \bibnamefont {Aissa}}, \bibinfo {author} {\bibfnamefont {S.}~\bibnamefont
  {Hage-Ali}}, \bibinfo {author} {\bibfnamefont {D.}~\bibnamefont {Lacour}},
  \bibinfo {author} {\bibfnamefont {F.}~\bibnamefont {Sarry}},\ and\ \bibinfo
  {author} {\bibfnamefont {O.}~\bibnamefont {Boumatar}},\ }\href
  {https://doi.org/10.1016/j.sna.2015.10.031} {\bibfield  {journal} {\bibinfo
  {journal} {Sensors and Actuators A: Physical}\ }\textbf {\bibinfo {volume}
  {240}},\ \bibinfo {pages} {41} (\bibinfo {year} {2016})}\BibitemShut
  {NoStop}%
\bibitem [{\citenamefont {Thevenard}\ \emph {et~al.}(2016)\citenamefont
  {Thevenard}, \citenamefont {Camara}, \citenamefont {Majrab}, \citenamefont
  {Bernard}, \citenamefont {Rovillain}, \citenamefont {Lema{\^{i}}tre},
  \citenamefont {Gourdon},\ and\ \citenamefont {Duquesne}}]{Thevenard2016}%
  \BibitemOpen
  \bibfield  {author} {\bibinfo {author} {\bibfnamefont {L.}~\bibnamefont
  {Thevenard}}, \bibinfo {author} {\bibfnamefont {I.~S.}\ \bibnamefont
  {Camara}}, \bibinfo {author} {\bibfnamefont {S.}~\bibnamefont {Majrab}},
  \bibinfo {author} {\bibfnamefont {M.}~\bibnamefont {Bernard}}, \bibinfo
  {author} {\bibfnamefont {P.}~\bibnamefont {Rovillain}}, \bibinfo {author}
  {\bibfnamefont {A.}~\bibnamefont {Lema{\^{i}}tre}}, \bibinfo {author}
  {\bibfnamefont {C.}~\bibnamefont {Gourdon}},\ and\ \bibinfo {author}
  {\bibfnamefont {J.-Y.}\ \bibnamefont {Duquesne}},\ }\href
  {https://doi.org/10.1103/PhysRevB.93.134430} {\bibfield  {journal} {\bibinfo
  {journal} {Physical Review B}\ }\textbf {\bibinfo {volume} {93}},\ \bibinfo
  {pages} {134430} (\bibinfo {year} {2016})}\BibitemShut {NoStop}%
\bibitem [{\citenamefont {Xu}\ \emph {et~al.}(2020)\citenamefont {Xu},
  \citenamefont {Yamamoto}, \citenamefont {Puebla}, \citenamefont {Baumgaertl},
  \citenamefont {Rana}, \citenamefont {Miura}, \citenamefont {Takahashi},
  \citenamefont {Grundler}, \citenamefont {Maekawa},\ and\ \citenamefont
  {Otani}}]{Xu2020}%
  \BibitemOpen
  \bibfield  {author} {\bibinfo {author} {\bibfnamefont {M.}~\bibnamefont
  {Xu}}, \bibinfo {author} {\bibfnamefont {K.}~\bibnamefont {Yamamoto}},
  \bibinfo {author} {\bibfnamefont {J.}~\bibnamefont {Puebla}}, \bibinfo
  {author} {\bibfnamefont {K.}~\bibnamefont {Baumgaertl}}, \bibinfo {author}
  {\bibfnamefont {B.}~\bibnamefont {Rana}}, \bibinfo {author} {\bibfnamefont
  {K.}~\bibnamefont {Miura}}, \bibinfo {author} {\bibfnamefont
  {H.}~\bibnamefont {Takahashi}}, \bibinfo {author} {\bibfnamefont
  {D.}~\bibnamefont {Grundler}}, \bibinfo {author} {\bibfnamefont
  {S.}~\bibnamefont {Maekawa}},\ and\ \bibinfo {author} {\bibfnamefont
  {Y.}~\bibnamefont {Otani}},\ }\href {https://doi.org/10.1126/sciadv.abb1724}
  {\bibfield  {journal} {\bibinfo  {journal} {Science Advances}\ }\textbf
  {\bibinfo {volume} {6}},\ \bibinfo {pages} {eabb1724} (\bibinfo {year}
  {2020})}\BibitemShut {NoStop}%
\bibitem [{\citenamefont {Babu}\ \emph {et~al.}(2021)\citenamefont {Babu},
  \citenamefont {Trzaskowska}, \citenamefont {Graczyk}, \citenamefont
  {Centa{\l}a}, \citenamefont {Mieszczak}, \citenamefont {G{\l}owi{\'{n}}ski},
  \citenamefont {Zdunek}, \citenamefont {Mielcarek},\ and\ \citenamefont
  {K{\l}os}}]{Babu2020}%
  \BibitemOpen
  \bibfield  {author} {\bibinfo {author} {\bibfnamefont {N.~K.~P.}\
  \bibnamefont {Babu}}, \bibinfo {author} {\bibfnamefont {A.}~\bibnamefont
  {Trzaskowska}}, \bibinfo {author} {\bibfnamefont {P.}~\bibnamefont
  {Graczyk}}, \bibinfo {author} {\bibfnamefont {G.}~\bibnamefont {Centa{\l}a}},
  \bibinfo {author} {\bibfnamefont {S.}~\bibnamefont {Mieszczak}}, \bibinfo
  {author} {\bibfnamefont {H.}~\bibnamefont {G{\l}owi{\'{n}}ski}}, \bibinfo
  {author} {\bibfnamefont {M.}~\bibnamefont {Zdunek}}, \bibinfo {author}
  {\bibfnamefont {S.}~\bibnamefont {Mielcarek}},\ and\ \bibinfo {author}
  {\bibfnamefont {J.~W.}\ \bibnamefont {K{\l}os}},\ }\href
  {https://doi.org/10.1021/acs.nanolett.0c03692} {\bibfield  {journal}
  {\bibinfo  {journal} {Nano Letters}\ }\textbf {\bibinfo {volume} {21}},\
  \bibinfo {pages} {946} (\bibinfo {year} {2021})}\BibitemShut {NoStop}%
\bibitem [{\citenamefont {K{\"{u}}{\ss}}\ \emph {et~al.}(2021)\citenamefont
  {K{\"{u}}{\ss}}, \citenamefont {Heigl}, \citenamefont {Flacke}, \citenamefont
  {Hefele}, \citenamefont {H{\"{o}}rner}, \citenamefont {Weiler}, \citenamefont
  {Albrecht},\ and\ \citenamefont {Wixforth}}]{Kus2021}%
  \BibitemOpen
  \bibfield  {author} {\bibinfo {author} {\bibfnamefont {M.}~\bibnamefont
  {K{\"{u}}{\ss}}}, \bibinfo {author} {\bibfnamefont {M.}~\bibnamefont
  {Heigl}}, \bibinfo {author} {\bibfnamefont {L.}~\bibnamefont {Flacke}},
  \bibinfo {author} {\bibfnamefont {A.}~\bibnamefont {Hefele}}, \bibinfo
  {author} {\bibfnamefont {A.}~\bibnamefont {H{\"{o}}rner}}, \bibinfo {author}
  {\bibfnamefont {M.}~\bibnamefont {Weiler}}, \bibinfo {author} {\bibfnamefont
  {M.}~\bibnamefont {Albrecht}},\ and\ \bibinfo {author} {\bibfnamefont
  {A.}~\bibnamefont {Wixforth}},\ }\href
  {https://doi.org/10.1103/PhysRevApplied.15.034046} {\bibfield  {journal}
  {\bibinfo  {journal} {Physical Review Applied}\ }\textbf {\bibinfo {volume}
  {15}},\ \bibinfo {pages} {034046} (\bibinfo {year} {2021})}\BibitemShut
  {NoStop}%
\bibitem [{\citenamefont {Verba}\ \emph {et~al.}(2019)\citenamefont {Verba},
  \citenamefont {Tiberkevich},\ and\ \citenamefont {Slavin}}]{Verba2019SA}%
  \BibitemOpen
  \bibfield  {author} {\bibinfo {author} {\bibfnamefont {R.}~\bibnamefont
  {Verba}}, \bibinfo {author} {\bibfnamefont {V.}~\bibnamefont {Tiberkevich}},\
  and\ \bibinfo {author} {\bibfnamefont {A.}~\bibnamefont {Slavin}},\ }\href
  {https://doi.org/10.1103/PhysRevApplied.12.054061} {\bibfield  {journal}
  {\bibinfo  {journal} {Physical Review Applied}\ }\textbf {\bibinfo {volume}
  {12}},\ \bibinfo {pages} {054061} (\bibinfo {year} {2019})}\BibitemShut
  {NoStop}%
\bibitem [{\citenamefont {Shah}\ \emph {et~al.}(2020)\citenamefont {Shah},
  \citenamefont {Bas}, \citenamefont {Lisenkov}, \citenamefont {Matyushov},
  \citenamefont {Sun},\ and\ \citenamefont {Page}}]{Shah2020}%
  \BibitemOpen
  \bibfield  {author} {\bibinfo {author} {\bibfnamefont {P.~J.}\ \bibnamefont
  {Shah}}, \bibinfo {author} {\bibfnamefont {D.~A.}\ \bibnamefont {Bas}},
  \bibinfo {author} {\bibfnamefont {I.}~\bibnamefont {Lisenkov}}, \bibinfo
  {author} {\bibfnamefont {A.}~\bibnamefont {Matyushov}}, \bibinfo {author}
  {\bibfnamefont {N.~X.}\ \bibnamefont {Sun}},\ and\ \bibinfo {author}
  {\bibfnamefont {M.~R.}\ \bibnamefont {Page}},\ }\href
  {https://doi.org/10.1126/sciadv.abc5648} {\bibfield  {journal} {\bibinfo
  {journal} {Science Advances}\ }\textbf {\bibinfo {volume} {6}},\ \bibinfo
  {pages} {eabc5648} (\bibinfo {year} {2020})}\BibitemShut {NoStop}%
\bibitem [{\citenamefont {Delsing}\ \emph {et~al.}(2019)\citenamefont
  {Delsing}, \citenamefont {Cleland}, \citenamefont {Schuetz}, \citenamefont
  {Kn{\"{o}}rzer}, \citenamefont {Giedke}, \citenamefont {Cirac}, \citenamefont
  {Srinivasan}, \citenamefont {Wu}, \citenamefont {Balram}, \citenamefont
  {Ba{\"{u}}erle}, \citenamefont {Meunier}, \citenamefont {Ford}, \citenamefont
  {Santos}, \citenamefont {Cerda-M{\'{e}}ndez}, \citenamefont {Wang},
  \citenamefont {Krenner}, \citenamefont {Nysten}, \citenamefont {Wei{\ss}},
  \citenamefont {Nash}, \citenamefont {Thevenard}, \citenamefont {Gourdon},
  \citenamefont {Rovillain}, \citenamefont {Marangolo}, \citenamefont
  {Duquesne}, \citenamefont {Fischerauer}, \citenamefont {Ruile}, \citenamefont
  {Reiner}, \citenamefont {Paschke}, \citenamefont {Denysenko}, \citenamefont
  {Volkmer}, \citenamefont {Wixforth}, \citenamefont {Bruus}, \citenamefont
  {Wiklund}, \citenamefont {Reboud}, \citenamefont {Cooper}, \citenamefont
  {Fu}, \citenamefont {Brugger}, \citenamefont {Rehfeldt},\ and\ \citenamefont
  {Westerhausen}}]{Delsing2019}%
  \BibitemOpen
  \bibfield  {author} {\bibinfo {author} {\bibfnamefont {P.}~\bibnamefont
  {Delsing}}, \bibinfo {author} {\bibfnamefont {A.~N.}\ \bibnamefont
  {Cleland}}, \bibinfo {author} {\bibfnamefont {M.~J.}\ \bibnamefont
  {Schuetz}}, \bibinfo {author} {\bibfnamefont {J.}~\bibnamefont
  {Kn{\"{o}}rzer}}, \bibinfo {author} {\bibfnamefont {G.}~\bibnamefont
  {Giedke}}, \bibinfo {author} {\bibfnamefont {J.~I.}\ \bibnamefont {Cirac}},
  \bibinfo {author} {\bibfnamefont {K.}~\bibnamefont {Srinivasan}}, \bibinfo
  {author} {\bibfnamefont {M.}~\bibnamefont {Wu}}, \bibinfo {author}
  {\bibfnamefont {K.~C.}\ \bibnamefont {Balram}}, \bibinfo {author}
  {\bibfnamefont {C.}~\bibnamefont {Ba{\"{u}}erle}}, \bibinfo {author}
  {\bibfnamefont {T.}~\bibnamefont {Meunier}}, \bibinfo {author} {\bibfnamefont
  {C.~J.}\ \bibnamefont {Ford}}, \bibinfo {author} {\bibfnamefont {P.~V.}\
  \bibnamefont {Santos}}, \bibinfo {author} {\bibfnamefont {E.}~\bibnamefont
  {Cerda-M{\'{e}}ndez}}, \bibinfo {author} {\bibfnamefont {H.}~\bibnamefont
  {Wang}}, \bibinfo {author} {\bibfnamefont {H.~J.}\ \bibnamefont {Krenner}},
  \bibinfo {author} {\bibfnamefont {E.~D.}\ \bibnamefont {Nysten}}, \bibinfo
  {author} {\bibfnamefont {M.}~\bibnamefont {Wei{\ss}}}, \bibinfo {author}
  {\bibfnamefont {G.~R.}\ \bibnamefont {Nash}}, \bibinfo {author}
  {\bibfnamefont {L.}~\bibnamefont {Thevenard}}, \bibinfo {author}
  {\bibfnamefont {C.}~\bibnamefont {Gourdon}}, \bibinfo {author} {\bibfnamefont
  {P.}~\bibnamefont {Rovillain}}, \bibinfo {author} {\bibfnamefont
  {M.}~\bibnamefont {Marangolo}}, \bibinfo {author} {\bibfnamefont {J.~Y.}\
  \bibnamefont {Duquesne}}, \bibinfo {author} {\bibfnamefont {G.}~\bibnamefont
  {Fischerauer}}, \bibinfo {author} {\bibfnamefont {W.}~\bibnamefont {Ruile}},
  \bibinfo {author} {\bibfnamefont {A.}~\bibnamefont {Reiner}}, \bibinfo
  {author} {\bibfnamefont {B.}~\bibnamefont {Paschke}}, \bibinfo {author}
  {\bibfnamefont {D.}~\bibnamefont {Denysenko}}, \bibinfo {author}
  {\bibfnamefont {D.}~\bibnamefont {Volkmer}}, \bibinfo {author} {\bibfnamefont
  {A.}~\bibnamefont {Wixforth}}, \bibinfo {author} {\bibfnamefont
  {H.}~\bibnamefont {Bruus}}, \bibinfo {author} {\bibfnamefont
  {M.}~\bibnamefont {Wiklund}}, \bibinfo {author} {\bibfnamefont
  {J.}~\bibnamefont {Reboud}}, \bibinfo {author} {\bibfnamefont {J.~M.}\
  \bibnamefont {Cooper}}, \bibinfo {author} {\bibfnamefont {Y.~Q.}\
  \bibnamefont {Fu}}, \bibinfo {author} {\bibfnamefont {M.~S.}\ \bibnamefont
  {Brugger}}, \bibinfo {author} {\bibfnamefont {F.}~\bibnamefont {Rehfeldt}},\
  and\ \bibinfo {author} {\bibfnamefont {C.}~\bibnamefont {Westerhausen}},\
  }\href {https://doi.org/10.1088/1361-6463/ab1b04} {\bibfield  {journal}
  {\bibinfo  {journal} {Journal of Physics D: Applied Physics}\ }\textbf
  {\bibinfo {volume} {52}},\ \bibinfo {pages} {353001} (\bibinfo {year}
  {2019})}\BibitemShut {NoStop}%
\bibitem [{\citenamefont {Caliendo}\ and\ \citenamefont
  {Hamidullah}(2019)}]{Han2019}%
  \BibitemOpen
  \bibfield  {author} {\bibinfo {author} {\bibfnamefont {C.}~\bibnamefont
  {Caliendo}}\ and\ \bibinfo {author} {\bibfnamefont {M.}~\bibnamefont
  {Hamidullah}},\ }\href {https://doi.org/10.1088/1361-6463/aafd0b} {\bibfield
  {journal} {\bibinfo  {journal} {Journal of Physics D: Applied Physics}\
  }\textbf {\bibinfo {volume} {52}},\ \bibinfo {pages} {153001} (\bibinfo
  {year} {2019})}\BibitemShut {NoStop}%
\bibitem [{\citenamefont {Weiler}\ \emph {et~al.}(2011)\citenamefont {Weiler},
  \citenamefont {Dreher}, \citenamefont {Heeg}, \citenamefont {Huebl},
  \citenamefont {Gross}, \citenamefont {Brandt},\ and\ \citenamefont
  {Goennenwein}}]{Weiler2011}%
  \BibitemOpen
  \bibfield  {author} {\bibinfo {author} {\bibfnamefont {M.}~\bibnamefont
  {Weiler}}, \bibinfo {author} {\bibfnamefont {L.}~\bibnamefont {Dreher}},
  \bibinfo {author} {\bibfnamefont {C.}~\bibnamefont {Heeg}}, \bibinfo {author}
  {\bibfnamefont {H.}~\bibnamefont {Huebl}}, \bibinfo {author} {\bibfnamefont
  {R.}~\bibnamefont {Gross}}, \bibinfo {author} {\bibfnamefont {M.~S.}\
  \bibnamefont {Brandt}},\ and\ \bibinfo {author} {\bibfnamefont {S.~T.~B.}\
  \bibnamefont {Goennenwein}},\ }\href
  {https://doi.org/10.1103/PhysRevLett.106.117601} {\bibfield  {journal}
  {\bibinfo  {journal} {Physical Review Letters}\ }\textbf {\bibinfo {volume}
  {106}},\ \bibinfo {pages} {117601} (\bibinfo {year} {2011})}\BibitemShut
  {NoStop}%
\bibitem [{\citenamefont {Yahagi}\ \emph {et~al.}(2017)\citenamefont {Yahagi},
  \citenamefont {Berk}, \citenamefont {Hebler}, \citenamefont {Dhuey},
  \citenamefont {Cabrini}, \citenamefont {Albrecht},\ and\ \citenamefont
  {Schmidt}}]{Yahagi2017}%
  \BibitemOpen
  \bibfield  {author} {\bibinfo {author} {\bibfnamefont {Y.}~\bibnamefont
  {Yahagi}}, \bibinfo {author} {\bibfnamefont {C.}~\bibnamefont {Berk}},
  \bibinfo {author} {\bibfnamefont {B.}~\bibnamefont {Hebler}}, \bibinfo
  {author} {\bibfnamefont {S.}~\bibnamefont {Dhuey}}, \bibinfo {author}
  {\bibfnamefont {S.}~\bibnamefont {Cabrini}}, \bibinfo {author} {\bibfnamefont
  {M.}~\bibnamefont {Albrecht}},\ and\ \bibinfo {author} {\bibfnamefont
  {H.}~\bibnamefont {Schmidt}},\ }\href
  {https://doi.org/10.1088/1361-6463/aa6472} {\bibfield  {journal} {\bibinfo
  {journal} {Journal of Physics D: Applied Physics}\ }\textbf {\bibinfo
  {volume} {50}},\ \bibinfo {pages} {17LT01} (\bibinfo {year}
  {2017})}\BibitemShut {NoStop}%
\bibitem [{\citenamefont {Kuszewski}\ \emph {et~al.}(2018)\citenamefont
  {Kuszewski}, \citenamefont {Duquesne}, \citenamefont {Becerra}, \citenamefont
  {Lema{\^{i}}tre}, \citenamefont {Vincent}, \citenamefont {Majrab},
  \citenamefont {Margaillan}, \citenamefont {Gourdon},\ and\ \citenamefont
  {Thevenard}}]{Kuszewski2018}%
  \BibitemOpen
  \bibfield  {author} {\bibinfo {author} {\bibfnamefont {P.}~\bibnamefont
  {Kuszewski}}, \bibinfo {author} {\bibfnamefont {J.~Y.}\ \bibnamefont
  {Duquesne}}, \bibinfo {author} {\bibfnamefont {L.}~\bibnamefont {Becerra}},
  \bibinfo {author} {\bibfnamefont {A.}~\bibnamefont {Lema{\^{i}}tre}},
  \bibinfo {author} {\bibfnamefont {S.}~\bibnamefont {Vincent}}, \bibinfo
  {author} {\bibfnamefont {S.}~\bibnamefont {Majrab}}, \bibinfo {author}
  {\bibfnamefont {F.}~\bibnamefont {Margaillan}}, \bibinfo {author}
  {\bibfnamefont {C.}~\bibnamefont {Gourdon}},\ and\ \bibinfo {author}
  {\bibfnamefont {L.}~\bibnamefont {Thevenard}},\ }\href
  {https://doi.org/10.1103/PhysRevApplied.10.034036} {\bibfield  {journal}
  {\bibinfo  {journal} {Physical Review Applied}\ }\textbf {\bibinfo {volume}
  {10}},\ \bibinfo {pages} {034036} (\bibinfo {year} {2018})}\BibitemShut
  {NoStop}%
\bibitem [{\citenamefont {K{\"{u}}{\ss}}\ \emph {et~al.}(2020)\citenamefont
  {K{\"{u}}{\ss}}, \citenamefont {Heigl}, \citenamefont {Flacke}, \citenamefont
  {H{\"{o}}rner}, \citenamefont {Weiler}, \citenamefont {Albrecht},\ and\
  \citenamefont {Wixforth}}]{Ku2020}%
  \BibitemOpen
  \bibfield  {author} {\bibinfo {author} {\bibfnamefont {M.}~\bibnamefont
  {K{\"{u}}{\ss}}}, \bibinfo {author} {\bibfnamefont {M.}~\bibnamefont
  {Heigl}}, \bibinfo {author} {\bibfnamefont {L.}~\bibnamefont {Flacke}},
  \bibinfo {author} {\bibfnamefont {A.}~\bibnamefont {H{\"{o}}rner}}, \bibinfo
  {author} {\bibfnamefont {M.}~\bibnamefont {Weiler}}, \bibinfo {author}
  {\bibfnamefont {M.}~\bibnamefont {Albrecht}},\ and\ \bibinfo {author}
  {\bibfnamefont {A.}~\bibnamefont {Wixforth}},\ }\href
  {https://doi.org/10.1103/PhysRevLett.125.217203} {\bibfield  {journal}
  {\bibinfo  {journal} {Physical Review Letters}\ }\textbf {\bibinfo {volume}
  {125}},\ \bibinfo {pages} {217203} (\bibinfo {year} {2020})}\BibitemShut
  {NoStop}%
\bibitem [{\citenamefont {Geilen}\ \emph {et~al.}(2022)\citenamefont {Geilen},
  \citenamefont {Nicoloiu}, \citenamefont {Narducci}, \citenamefont {Mohseni},
  \citenamefont {Bechberger}, \citenamefont {Ender}, \citenamefont {Ciubotaru},
  \citenamefont {Hillebrands}, \citenamefont {Müller}, \citenamefont
  {Adelmann},\ and\ \citenamefont {Pirro}}]{geilen2021fully}%
  \BibitemOpen
  \bibfield  {author} {\bibinfo {author} {\bibfnamefont {M.}~\bibnamefont
  {Geilen}}, \bibinfo {author} {\bibfnamefont {A.}~\bibnamefont {Nicoloiu}},
  \bibinfo {author} {\bibfnamefont {D.}~\bibnamefont {Narducci}}, \bibinfo
  {author} {\bibfnamefont {M.}~\bibnamefont {Mohseni}}, \bibinfo {author}
  {\bibfnamefont {M.}~\bibnamefont {Bechberger}}, \bibinfo {author}
  {\bibfnamefont {M.}~\bibnamefont {Ender}}, \bibinfo {author} {\bibfnamefont
  {F.}~\bibnamefont {Ciubotaru}}, \bibinfo {author} {\bibfnamefont
  {B.}~\bibnamefont {Hillebrands}}, \bibinfo {author} {\bibfnamefont
  {A.}~\bibnamefont {Müller}}, \bibinfo {author} {\bibfnamefont
  {C.}~\bibnamefont {Adelmann}},\ and\ \bibinfo {author} {\bibfnamefont
  {P.}~\bibnamefont {Pirro}},\ }\href {https://doi.org/10.1063/5.0088924}
  {\bibfield  {journal} {\bibinfo  {journal} {Applied Physics Letters}\
  }\textbf {\bibinfo {volume} {120}},\ \bibinfo {pages} {242404} (\bibinfo
  {year} {2022})}\BibitemShut {NoStop}%
\bibitem [{\citenamefont {Mahmoud}\ \emph {et~al.}(2020)\citenamefont
  {Mahmoud}, \citenamefont {Ciubotaru}, \citenamefont {Vanderveken},
  \citenamefont {Chumak}, \citenamefont {Hamdioui}, \citenamefont {Adelmann},\
  and\ \citenamefont {Cotofana}}]{Mahmoud2020}%
  \BibitemOpen
  \bibfield  {author} {\bibinfo {author} {\bibfnamefont {A.}~\bibnamefont
  {Mahmoud}}, \bibinfo {author} {\bibfnamefont {F.}~\bibnamefont {Ciubotaru}},
  \bibinfo {author} {\bibfnamefont {F.}~\bibnamefont {Vanderveken}}, \bibinfo
  {author} {\bibfnamefont {A.~V.}\ \bibnamefont {Chumak}}, \bibinfo {author}
  {\bibfnamefont {S.}~\bibnamefont {Hamdioui}}, \bibinfo {author}
  {\bibfnamefont {C.}~\bibnamefont {Adelmann}},\ and\ \bibinfo {author}
  {\bibfnamefont {S.}~\bibnamefont {Cotofana}},\ }\href
  {https://aip.scitation.org/doi/abs/10.1063/5.0019328} {\bibfield  {journal}
  {\bibinfo  {journal} {Journal of Applied Physics}\ }\textbf {\bibinfo
  {volume} {128}},\ \bibinfo {pages} {161101} (\bibinfo {year}
  {2020})}\BibitemShut {NoStop}%
\bibitem [{\citenamefont {Pirro}\ \emph {et~al.}(2021)\citenamefont {Pirro},
  \citenamefont {Vasyuchka}, \citenamefont {Serga},\ and\ \citenamefont
  {Hillebrands}}]{Pirro2021}%
  \BibitemOpen
  \bibfield  {author} {\bibinfo {author} {\bibfnamefont {P.}~\bibnamefont
  {Pirro}}, \bibinfo {author} {\bibfnamefont {V.}~\bibnamefont {Vasyuchka}},
  \bibinfo {author} {\bibfnamefont {A.~A.}\ \bibnamefont {Serga}},\ and\
  \bibinfo {author} {\bibfnamefont {B.}~\bibnamefont {Hillebrands}},\ }\href
  {https://doi.org/10.1038/s41578-021-00332-w} {\bibfield  {journal} {\bibinfo
  {journal} {Nature Reviews Materials}\ }\textbf {\bibinfo {volume} {6}},\
  \bibinfo {pages} {227601} (\bibinfo {year} {2021})}\BibitemShut {NoStop}%
\bibitem [{\citenamefont {Pirro}\ \emph {et~al.}(2014)\citenamefont {Pirro},
  \citenamefont {Sebastian}, \citenamefont {Br{\"{a}}cher}, \citenamefont
  {Serga}, \citenamefont {Kubota}, \citenamefont {Naganuma}, \citenamefont
  {Oogane}, \citenamefont {Ando},\ and\ \citenamefont
  {Hillebrands}}]{Pirro2014}%
  \BibitemOpen
  \bibfield  {author} {\bibinfo {author} {\bibfnamefont {P.}~\bibnamefont
  {Pirro}}, \bibinfo {author} {\bibfnamefont {T.}~\bibnamefont {Sebastian}},
  \bibinfo {author} {\bibfnamefont {T.}~\bibnamefont {Br{\"{a}}cher}}, \bibinfo
  {author} {\bibfnamefont {A.~A.}\ \bibnamefont {Serga}}, \bibinfo {author}
  {\bibfnamefont {T.}~\bibnamefont {Kubota}}, \bibinfo {author} {\bibfnamefont
  {H.}~\bibnamefont {Naganuma}}, \bibinfo {author} {\bibfnamefont
  {M.}~\bibnamefont {Oogane}}, \bibinfo {author} {\bibfnamefont
  {Y.}~\bibnamefont {Ando}},\ and\ \bibinfo {author} {\bibfnamefont
  {B.}~\bibnamefont {Hillebrands}},\ }\href
  {https://doi.org/10.1103/PhysRevLett.113.227601} {\bibfield  {journal}
  {\bibinfo  {journal} {Physical Review Letters}\ }\textbf {\bibinfo {volume}
  {113}},\ \bibinfo {pages} {227601} (\bibinfo {year} {2014})}\BibitemShut
  {NoStop}%
\bibitem [{\citenamefont {Wang}\ \emph {et~al.}(2020)\citenamefont {Wang},
  \citenamefont {Hamadeh}, \citenamefont {Verba}, \citenamefont {Lomakin},
  \citenamefont {Mohseni}, \citenamefont {Hillebrands}, \citenamefont
  {Chumak},\ and\ \citenamefont {Pirro}}]{Wang2020}%
  \BibitemOpen
  \bibfield  {author} {\bibinfo {author} {\bibfnamefont {Q.}~\bibnamefont
  {Wang}}, \bibinfo {author} {\bibfnamefont {A.}~\bibnamefont {Hamadeh}},
  \bibinfo {author} {\bibfnamefont {R.}~\bibnamefont {Verba}}, \bibinfo
  {author} {\bibfnamefont {V.}~\bibnamefont {Lomakin}}, \bibinfo {author}
  {\bibfnamefont {M.}~\bibnamefont {Mohseni}}, \bibinfo {author} {\bibfnamefont
  {B.}~\bibnamefont {Hillebrands}}, \bibinfo {author} {\bibfnamefont {A.~V.}\
  \bibnamefont {Chumak}},\ and\ \bibinfo {author} {\bibfnamefont
  {P.}~\bibnamefont {Pirro}},\ }\href
  {https://doi.org/10.1038/s41524-020-00465-6} {\bibfield  {journal} {\bibinfo
  {journal} {npj Computational Materials}\ }\textbf {\bibinfo {volume} {6}},\
  \bibinfo {pages} {192} (\bibinfo {year} {2020})}\BibitemShut {NoStop}%
\bibitem [{\citenamefont {Schultheiss}\ \emph {et~al.}(2012)\citenamefont
  {Schultheiss}, \citenamefont {Vogt},\ and\ \citenamefont
  {Hillebrands}}]{Schultheiss2012}%
  \BibitemOpen
  \bibfield  {author} {\bibinfo {author} {\bibfnamefont {H.}~\bibnamefont
  {Schultheiss}}, \bibinfo {author} {\bibfnamefont {K.}~\bibnamefont {Vogt}},\
  and\ \bibinfo {author} {\bibfnamefont {B.}~\bibnamefont {Hillebrands}},\
  }\href {https://doi.org/10.1103/PhysRevB.86.054414} {\bibfield  {journal}
  {\bibinfo  {journal} {Physical Review B}\ }\textbf {\bibinfo {volume} {86}},\
  \bibinfo {pages} {054414} (\bibinfo {year} {2012})}\BibitemShut {NoStop}%
\bibitem [{\citenamefont {Krivosik}\ and\ \citenamefont
  {Patton}(2010)}]{Krivosik2010}%
  \BibitemOpen
  \bibfield  {author} {\bibinfo {author} {\bibfnamefont {P.}~\bibnamefont
  {Krivosik}}\ and\ \bibinfo {author} {\bibfnamefont {C.~E.}\ \bibnamefont
  {Patton}},\ }\href {https://doi.org/10.1103/PhysRevB.82.184428} {\bibfield
  {journal} {\bibinfo  {journal} {Physical Review B}\ }\textbf {\bibinfo
  {volume} {82}},\ \bibinfo {pages} {184428} (\bibinfo {year}
  {2010})}\BibitemShut {NoStop}%
\bibitem [{\citenamefont {Papp}\ \emph {et~al.}(2021)\citenamefont {Papp},
  \citenamefont {Porod},\ and\ \citenamefont {Csaba}}]{Papp2021}%
  \BibitemOpen
  \bibfield  {author} {\bibinfo {author} {\bibfnamefont {{\'{A}}.}~\bibnamefont
  {Papp}}, \bibinfo {author} {\bibfnamefont {W.}~\bibnamefont {Porod}},\ and\
  \bibinfo {author} {\bibfnamefont {G.}~\bibnamefont {Csaba}},\ }\href
  {https://doi.org/10.1038/s41467-021-26711-z} {\bibfield  {journal} {\bibinfo
  {journal} {Nature Communications}\ }\textbf {\bibinfo {volume} {12}},\
  \bibinfo {pages} {6422} (\bibinfo {year} {2021})}\BibitemShut {NoStop}%
\bibitem [{\citenamefont {Kraimia}\ \emph {et~al.}(2020)\citenamefont
  {Kraimia}, \citenamefont {Kuszewski}, \citenamefont {Duquesne}, \citenamefont
  {Lema{\^{i}}tre}, \citenamefont {Margaillan}, \citenamefont {Gourdon},\ and\
  \citenamefont {Thevenard}}]{Kraimia2020}%
  \BibitemOpen
  \bibfield  {author} {\bibinfo {author} {\bibfnamefont {M.}~\bibnamefont
  {Kraimia}}, \bibinfo {author} {\bibfnamefont {P.}~\bibnamefont {Kuszewski}},
  \bibinfo {author} {\bibfnamefont {J.-Y.}\ \bibnamefont {Duquesne}}, \bibinfo
  {author} {\bibfnamefont {A.}~\bibnamefont {Lema{\^{i}}tre}}, \bibinfo
  {author} {\bibfnamefont {F.}~\bibnamefont {Margaillan}}, \bibinfo {author}
  {\bibfnamefont {C.}~\bibnamefont {Gourdon}},\ and\ \bibinfo {author}
  {\bibfnamefont {L.}~\bibnamefont {Thevenard}},\ }\href
  {https://doi.org/10.1103/PhysRevB.101.144425} {\bibfield  {journal} {\bibinfo
   {journal} {Physical Review B}\ }\textbf {\bibinfo {volume} {101}},\ \bibinfo
  {pages} {144425} (\bibinfo {year} {2020})}\BibitemShut {NoStop}%
\bibitem [{\citenamefont {Lisenkov}\ \emph {et~al.}(2019)\citenamefont
  {Lisenkov}, \citenamefont {Jander},\ and\ \citenamefont
  {Dhagat}}]{Lisenkov2019}%
  \BibitemOpen
  \bibfield  {author} {\bibinfo {author} {\bibfnamefont {I.}~\bibnamefont
  {Lisenkov}}, \bibinfo {author} {\bibfnamefont {A.}~\bibnamefont {Jander}},\
  and\ \bibinfo {author} {\bibfnamefont {P.}~\bibnamefont {Dhagat}},\ }\href
  {https://doi.org/10.1103/PhysRevB.99.184433} {\bibfield  {journal} {\bibinfo
  {journal} {Physical Review B}\ }\textbf {\bibinfo {volume} {99}},\ \bibinfo
  {pages} {184433} (\bibinfo {year} {2019})}\BibitemShut {NoStop}%
\bibitem [{\citenamefont {Zhang}\ \emph {et~al.}(2020)\citenamefont {Zhang},
  \citenamefont {Bauer},\ and\ \citenamefont {Yu}}]{Zhang2020}%
  \BibitemOpen
  \bibfield  {author} {\bibinfo {author} {\bibfnamefont {X.}~\bibnamefont
  {Zhang}}, \bibinfo {author} {\bibfnamefont {G.~E.}\ \bibnamefont {Bauer}},\
  and\ \bibinfo {author} {\bibfnamefont {T.}~\bibnamefont {Yu}},\ }\href
  {https://doi.org/10.1103/PhysRevLett.125.077203} {\bibfield  {journal}
  {\bibinfo  {journal} {Physical Review Letters}\ }\textbf {\bibinfo {volume}
  {125}},\ \bibinfo {pages} {077203} (\bibinfo {year} {2020})}\BibitemShut
  {NoStop}%
\bibitem [{\citenamefont {Geilen}\ \emph {et~al.}(2020)\citenamefont {Geilen},
  \citenamefont {Kohl}, \citenamefont {Nicoloiu}, \citenamefont {M{\"{u}}ller},
  \citenamefont {Hillebrands},\ and\ \citenamefont {Pirro}}]{Geilen2020}%
  \BibitemOpen
  \bibfield  {author} {\bibinfo {author} {\bibfnamefont {M.}~\bibnamefont
  {Geilen}}, \bibinfo {author} {\bibfnamefont {F.}~\bibnamefont {Kohl}},
  \bibinfo {author} {\bibfnamefont {A.}~\bibnamefont {Nicoloiu}}, \bibinfo
  {author} {\bibfnamefont {A.}~\bibnamefont {M{\"{u}}ller}}, \bibinfo {author}
  {\bibfnamefont {B.}~\bibnamefont {Hillebrands}},\ and\ \bibinfo {author}
  {\bibfnamefont {P.}~\bibnamefont {Pirro}},\ }\href
  {https://doi.org/10.1063/5.0029308} {\bibfield  {journal} {\bibinfo
  {journal} {Applied Physics Letters}\ }\textbf {\bibinfo {volume} {117}},\
  \bibinfo {pages} {213501} (\bibinfo {year} {2020})}\BibitemShut {NoStop}%
\bibitem [{\citenamefont {Sebastian}\ \emph {et~al.}(2015)\citenamefont
  {Sebastian}, \citenamefont {Schultheiss}, \citenamefont {Obry}, \citenamefont
  {Hillebrands},\ and\ \citenamefont {Schultheiss}}]{Sebastian2015}%
  \BibitemOpen
  \bibfield  {author} {\bibinfo {author} {\bibfnamefont {T.}~\bibnamefont
  {Sebastian}}, \bibinfo {author} {\bibfnamefont {K.}~\bibnamefont
  {Schultheiss}}, \bibinfo {author} {\bibfnamefont {B.}~\bibnamefont {Obry}},
  \bibinfo {author} {\bibfnamefont {B.}~\bibnamefont {Hillebrands}},\ and\
  \bibinfo {author} {\bibfnamefont {H.}~\bibnamefont {Schultheiss}},\ }\href
  {https://doi.org/10.3389/fphy.2015.00035} {\bibfield  {journal} {\bibinfo
  {journal} {Frontiers in Physics}\ }\textbf {\bibinfo {volume} {3}},\ \bibinfo
  {pages} {35} (\bibinfo {year} {2015})}\BibitemShut {NoStop}%
\bibitem [{\citenamefont {Kargar}\ and\ \citenamefont
  {Balandin}(2021)}]{Kargar2021}%
  \BibitemOpen
  \bibfield  {author} {\bibinfo {author} {\bibfnamefont {F.}~\bibnamefont
  {Kargar}}\ and\ \bibinfo {author} {\bibfnamefont {A.~A.}\ \bibnamefont
  {Balandin}},\ }\href {https://doi.org/10.1038/s41566-021-00836-5} {\bibfield
  {journal} {\bibinfo  {journal} {Nature Photonics}\ }\textbf {\bibinfo
  {volume} {15}},\ \bibinfo {pages} {720–73} (\bibinfo {year}
  {2021})}\BibitemShut {NoStop}%
\bibitem [{\citenamefont {Gurevich}\ and\ \citenamefont
  {Melkov}(1996)}]{Gurevich1996}%
  \BibitemOpen
  \bibfield  {author} {\bibinfo {author} {\bibfnamefont {A.}~\bibnamefont
  {Gurevich}}\ and\ \bibinfo {author} {\bibfnamefont {G.~A.}\ \bibnamefont
  {Melkov}},\ }\href@noop {} {\emph {\bibinfo {title} {Magnetization
  Oscillations and Waves}}}\ (\bibinfo  {publisher} {CRC Press},\ \bibinfo
  {year} {1996})\BibitemShut {NoStop}%
\bibitem [{\citenamefont {Prabhakar}\ and\ \citenamefont
  {Stancil}(2009)}]{Stancil2009}%
  \BibitemOpen
  \bibfield  {author} {\bibinfo {author} {\bibfnamefont {A.}~\bibnamefont
  {Prabhakar}}\ and\ \bibinfo {author} {\bibfnamefont {D.~D.}\ \bibnamefont
  {Stancil}},\ }\href {https://doi.org/10.1007/978-0-387-77865-5} {\emph
  {\bibinfo {title} {{Spin Waves: Theory and Applications}}}}\ (\bibinfo
  {publisher} {Springer New York, NY},\ \bibinfo {year} {2009})\BibitemShut
  {NoStop}%
\bibitem [{\citenamefont {Mayer}(2008)}]{Mayer2008}%
  \BibitemOpen
  \bibfield  {author} {\bibinfo {author} {\bibfnamefont {A.~P.}\ \bibnamefont
  {Mayer}},\ }\href {https://doi.org/10.1016/j.ultras.2008.06.009} {\bibfield
  {journal} {\bibinfo  {journal} {Ultrasonics}\ }\textbf {\bibinfo {volume}
  {48}},\ \bibinfo {pages} {478} (\bibinfo {year} {2008})}\BibitemShut
  {NoStop}%
\bibitem [{\citenamefont {Mayer}(1995)}]{Mayer1995}%
  \BibitemOpen
  \bibfield  {author} {\bibinfo {author} {\bibfnamefont {A.~P.}\ \bibnamefont
  {Mayer}},\ }\href {https://doi.org/10.1016/0370-1573(94)00088-K} {\bibfield
  {journal} {\bibinfo  {journal} {Physics Reports}\ }\textbf {\bibinfo {volume}
  {256}},\ \bibinfo {pages} {237} (\bibinfo {year} {1995})}\BibitemShut
  {NoStop}%
\bibitem [{\citenamefont {R\"uckriegel}\ \emph {et~al.}(2014)\citenamefont
  {R\"uckriegel}, \citenamefont {Kopietz}, \citenamefont {Bozhko},
  \citenamefont {Serga},\ and\ \citenamefont {Hillebrands}}]{Ruckspiegel2014}%
  \BibitemOpen
  \bibfield  {author} {\bibinfo {author} {\bibfnamefont {A.}~\bibnamefont
  {R\"uckriegel}}, \bibinfo {author} {\bibfnamefont {P.}~\bibnamefont
  {Kopietz}}, \bibinfo {author} {\bibfnamefont {D.~A.}\ \bibnamefont {Bozhko}},
  \bibinfo {author} {\bibfnamefont {A.~A.}\ \bibnamefont {Serga}},\ and\
  \bibinfo {author} {\bibfnamefont {B.}~\bibnamefont {Hillebrands}},\ }\href
  {https://doi.org/10.1103/PhysRevB.89.184413} {\bibfield  {journal} {\bibinfo
  {journal} {Phys. Rev. B}\ }\textbf {\bibinfo {volume} {89}},\ \bibinfo
  {pages} {184413} (\bibinfo {year} {2014})}\BibitemShut {NoStop}%
\bibitem [{\citenamefont {Vansteenkiste}\ \emph {et~al.}(2014)\citenamefont
  {Vansteenkiste}, \citenamefont {Leliaert}, \citenamefont {Dvornik},
  \citenamefont {Helsen}, \citenamefont {Garcia-Sanchez},\ and\ \citenamefont
  {{Van Waeyenberge}}}]{Vansteenkiste2014}%
  \BibitemOpen
  \bibfield  {author} {\bibinfo {author} {\bibfnamefont {A.}~\bibnamefont
  {Vansteenkiste}}, \bibinfo {author} {\bibfnamefont {J.}~\bibnamefont
  {Leliaert}}, \bibinfo {author} {\bibfnamefont {M.}~\bibnamefont {Dvornik}},
  \bibinfo {author} {\bibfnamefont {M.}~\bibnamefont {Helsen}}, \bibinfo
  {author} {\bibfnamefont {F.}~\bibnamefont {Garcia-Sanchez}},\ and\ \bibinfo
  {author} {\bibfnamefont {B.}~\bibnamefont {{Van Waeyenberge}}},\ }\href
  {https://doi.org/10.1063/1.4899186} {\bibfield  {journal} {\bibinfo
  {journal} {AIP Advances}\ }\textbf {\bibinfo {volume} {4}},\ \bibinfo {pages}
  {107133} (\bibinfo {year} {2014})}\BibitemShut {NoStop}%
\bibitem [{ait()}]{aithericon}%
  \BibitemOpen
  \href {https://aithericon.com} {\bibinfo {title}
  {aithericon.com}}\BibitemShut {NoStop}%
\bibitem [{Sup()}]{Supplement}%
  \BibitemOpen
  \href@noop {} {\bibinfo {title} {Supplement material}}\BibitemShut {NoStop}%
\bibitem [{\citenamefont {Br{\"{a}}cher}\ \emph {et~al.}(2017)\citenamefont
  {Br{\"{a}}cher}, \citenamefont {Pirro},\ and\ \citenamefont
  {Hillebrands}}]{Bracher2017}%
  \BibitemOpen
  \bibfield  {author} {\bibinfo {author} {\bibfnamefont {T.}~\bibnamefont
  {Br{\"{a}}cher}}, \bibinfo {author} {\bibfnamefont {P.}~\bibnamefont
  {Pirro}},\ and\ \bibinfo {author} {\bibfnamefont {B.}~\bibnamefont
  {Hillebrands}},\ }\href {https://doi.org/10.1016/j.physrep.2017.07.003}
  {\bibfield  {journal} {\bibinfo  {journal} {Physics Reports}\ }\textbf
  {\bibinfo {volume} {699}},\ \bibinfo {pages} {1} (\bibinfo {year}
  {2017})}\BibitemShut {NoStop}%
\bibitem [{\citenamefont {Verba}\ \emph {et~al.}(2017)\citenamefont {Verba},
  \citenamefont {Carpentieri}, \citenamefont {Finocchio}, \citenamefont
  {Tiberkevich},\ and\ \citenamefont {Slavin}}]{Verba2017}%
  \BibitemOpen
  \bibfield  {author} {\bibinfo {author} {\bibfnamefont {R.}~\bibnamefont
  {Verba}}, \bibinfo {author} {\bibfnamefont {M.}~\bibnamefont {Carpentieri}},
  \bibinfo {author} {\bibfnamefont {G.}~\bibnamefont {Finocchio}}, \bibinfo
  {author} {\bibfnamefont {V.}~\bibnamefont {Tiberkevich}},\ and\ \bibinfo
  {author} {\bibfnamefont {A.}~\bibnamefont {Slavin}},\ }\href
  {https://doi.org/10.1103/PhysRevApplied.7.064023} {\bibfield  {journal}
  {\bibinfo  {journal} {Physical Review Applied}\ }\textbf {\bibinfo {volume}
  {7}},\ \bibinfo {pages} {064023} (\bibinfo {year} {2017})}\BibitemShut
  {NoStop}%
\end{thebibliography}%


\begin{thebibliography}{7}%
\makeatletter
\providecommand \@ifxundefined [1]{%
 \@ifx{#1\undefined}
}%
\providecommand \@ifnum [1]{%
 \ifnum #1\expandafter \@firstoftwo
 \else \expandafter \@secondoftwo
 \fi
}%
\providecommand \@ifx [1]{%
 \ifx #1\expandafter \@firstoftwo
 \else \expandafter \@secondoftwo
 \fi
}%
\providecommand \natexlab [1]{#1}%
\providecommand \enquote  [1]{``#1''}%
\providecommand \bibnamefont  [1]{#1}%
\providecommand \bibfnamefont [1]{#1}%
\providecommand \citenamefont [1]{#1}%
\providecommand \href@noop [0]{\@secondoftwo}%
\providecommand \href [0]{\begingroup \@sanitize@url \@href}%
\providecommand \@href[1]{\@@startlink{#1}\@@href}%
\providecommand \@@href[1]{\endgroup#1\@@endlink}%
\providecommand \@sanitize@url [0]{\catcode `\\12\catcode `\$12\catcode
  `\&12\catcode `\#12\catcode `\^12\catcode `\_12\catcode `\%12\relax}%
\providecommand \@@startlink[1]{}%
\providecommand \@@endlink[0]{}%
\providecommand \url  [0]{\begingroup\@sanitize@url \@url }%
\providecommand \@url [1]{\endgroup\@href {#1}{\urlprefix }}%
\providecommand \urlprefix  [0]{URL }%
\providecommand \Eprint [0]{\href }%
\providecommand \doibase [0]{https://doi.org/}%
\providecommand \selectlanguage [0]{\@gobble}%
\providecommand \bibinfo  [0]{\@secondoftwo}%
\providecommand \bibfield  [0]{\@secondoftwo}%
\providecommand \translation [1]{[#1]}%
\providecommand \BibitemOpen [0]{}%
\providecommand \bibitemStop [0]{}%
\providecommand \bibitemNoStop [0]{.\EOS\space}%
\providecommand \EOS [0]{\spacefactor3000\relax}%
\providecommand \BibitemShut  [1]{\csname bibitem#1\endcsname}%
\let\auto@bib@innerbib\@empty
\bibitem [{\citenamefont {Vanderveken}\ \emph {et~al.}(2021)\citenamefont
  {Vanderveken}, \citenamefont {Mulkers}, \citenamefont {Leliaert},
  \citenamefont {{Van Waeyenberge}}, \citenamefont {Sor{\'{e}}e}, \citenamefont
  {Zografos}, \citenamefont {Ciubotaru},\ and\ \citenamefont
  {Adelmann}}]{Vanderveken2021}%
  \BibitemOpen
  \bibfield  {author} {\bibinfo {author} {\bibfnamefont {F.}~\bibnamefont
  {Vanderveken}}, \bibinfo {author} {\bibfnamefont {J.}~\bibnamefont
  {Mulkers}}, \bibinfo {author} {\bibfnamefont {J.}~\bibnamefont {Leliaert}},
  \bibinfo {author} {\bibfnamefont {B.}~\bibnamefont {{Van Waeyenberge}}},
  \bibinfo {author} {\bibfnamefont {B.}~\bibnamefont {Sor{\'{e}}e}}, \bibinfo
  {author} {\bibfnamefont {O.}~\bibnamefont {Zografos}}, \bibinfo {author}
  {\bibfnamefont {F.}~\bibnamefont {Ciubotaru}},\ and\ \bibinfo {author}
  {\bibfnamefont {C.}~\bibnamefont {Adelmann}},\ }\href
  {https://doi.org/10.1103/PhysRevB.103.054439} {\bibfield  {journal} {\bibinfo
   {journal} {Physical Review B}\ }\textbf {\bibinfo {volume} {103}},\ \bibinfo
  {pages} {054439} (\bibinfo {year} {2021})}\BibitemShut {NoStop}%
\bibitem [{\citenamefont {Geilen}\ \emph {et~al.}(2022)\citenamefont {Geilen},
  \citenamefont {Nicoloiu}, \citenamefont {Narducci}, \citenamefont {Mohseni},
  \citenamefont {Bechberger}, \citenamefont {Ender}, \citenamefont {Ciubotaru},
  \citenamefont {Hillebrands}, \citenamefont {Müller}, \citenamefont
  {Adelmann},\ and\ \citenamefont {Pirro}}]{geilen2021fully}%
  \BibitemOpen
  \bibfield  {author} {\bibinfo {author} {\bibfnamefont {M.}~\bibnamefont
  {Geilen}}, \bibinfo {author} {\bibfnamefont {A.}~\bibnamefont {Nicoloiu}},
  \bibinfo {author} {\bibfnamefont {D.}~\bibnamefont {Narducci}}, \bibinfo
  {author} {\bibfnamefont {M.}~\bibnamefont {Mohseni}}, \bibinfo {author}
  {\bibfnamefont {M.}~\bibnamefont {Bechberger}}, \bibinfo {author}
  {\bibfnamefont {M.}~\bibnamefont {Ender}}, \bibinfo {author} {\bibfnamefont
  {F.}~\bibnamefont {Ciubotaru}}, \bibinfo {author} {\bibfnamefont
  {B.}~\bibnamefont {Hillebrands}}, \bibinfo {author} {\bibfnamefont
  {A.}~\bibnamefont {Müller}}, \bibinfo {author} {\bibfnamefont
  {C.}~\bibnamefont {Adelmann}},\ and\ \bibinfo {author} {\bibfnamefont
  {P.}~\bibnamefont {Pirro}},\ }\href {https://doi.org/10.1063/5.0088924}
  {\bibfield  {journal} {\bibinfo  {journal} {Applied Physics Letters}\
  }\textbf {\bibinfo {volume} {120}},\ \bibinfo {pages} {242404} (\bibinfo
  {year} {2022})}\BibitemShut {NoStop}%
\bibitem [{\citenamefont {Lisenkov}\ \emph {et~al.}(2019)\citenamefont
  {Lisenkov}, \citenamefont {Jander},\ and\ \citenamefont
  {Dhagat}}]{Lisenkov2019}%
  \BibitemOpen
  \bibfield  {author} {\bibinfo {author} {\bibfnamefont {I.}~\bibnamefont
  {Lisenkov}}, \bibinfo {author} {\bibfnamefont {A.}~\bibnamefont {Jander}},\
  and\ \bibinfo {author} {\bibfnamefont {P.}~\bibnamefont {Dhagat}},\ }\href
  {https://doi.org/10.1103/PhysRevB.99.184433} {\bibfield  {journal} {\bibinfo
  {journal} {Physical Review B}\ }\textbf {\bibinfo {volume} {99}},\ \bibinfo
  {pages} {184433} (\bibinfo {year} {2019})}\BibitemShut {NoStop}%
\bibitem [{\citenamefont {Verba}\ \emph {et~al.}(2014)\citenamefont {Verba},
  \citenamefont {Tiberkevich}, \citenamefont {Krivorotov},\ and\ \citenamefont
  {Slavin}}]{Verba2014}%
  \BibitemOpen
  \bibfield  {author} {\bibinfo {author} {\bibfnamefont {R.}~\bibnamefont
  {Verba}}, \bibinfo {author} {\bibfnamefont {V.}~\bibnamefont {Tiberkevich}},
  \bibinfo {author} {\bibfnamefont {I.}~\bibnamefont {Krivorotov}},\ and\
  \bibinfo {author} {\bibfnamefont {A.}~\bibnamefont {Slavin}},\ }\href
  {https://doi.org/10.1103/PhysRevApplied.1.044006} {\bibfield  {journal}
  {\bibinfo  {journal} {Physical Review Applied}\ }\textbf {\bibinfo {volume}
  {1}},\ \bibinfo {pages} {044006} (\bibinfo {year} {2014})}\BibitemShut
  {NoStop}%
\bibitem [{\citenamefont {Verba}\ \emph {et~al.}(2021)\citenamefont {Verba},
  \citenamefont {K{\"{o}}rber}, \citenamefont {Schultheiss}, \citenamefont
  {Schultheiss}, \citenamefont {Tiberkevich},\ and\ \citenamefont
  {Slavin}}]{Verba2021}%
  \BibitemOpen
  \bibfield  {author} {\bibinfo {author} {\bibfnamefont {R.}~\bibnamefont
  {Verba}}, \bibinfo {author} {\bibfnamefont {L.}~\bibnamefont {K{\"{o}}rber}},
  \bibinfo {author} {\bibfnamefont {K.}~\bibnamefont {Schultheiss}}, \bibinfo
  {author} {\bibfnamefont {H.}~\bibnamefont {Schultheiss}}, \bibinfo {author}
  {\bibfnamefont {V.}~\bibnamefont {Tiberkevich}},\ and\ \bibinfo {author}
  {\bibfnamefont {A.}~\bibnamefont {Slavin}},\ }\href
  {https://doi.org/10.1103/PhysRevB.103.014413} {\bibfield  {journal} {\bibinfo
   {journal} {Physical Review B}\ }\textbf {\bibinfo {volume} {103}},\ \bibinfo
  {pages} {014413} (\bibinfo {year} {2021})}\BibitemShut {NoStop}%
\bibitem [{\citenamefont {Krivosik}\ and\ \citenamefont
  {Patton}(2010)}]{Krivosik2010}%
  \BibitemOpen
  \bibfield  {author} {\bibinfo {author} {\bibfnamefont {P.}~\bibnamefont
  {Krivosik}}\ and\ \bibinfo {author} {\bibfnamefont {C.~E.}\ \bibnamefont
  {Patton}},\ }\href {https://doi.org/10.1103/PhysRevB.82.184428} {\bibfield
  {journal} {\bibinfo  {journal} {Physical Review B}\ }\textbf {\bibinfo
  {volume} {82}},\ \bibinfo {pages} {184428} (\bibinfo {year}
  {2010})}\BibitemShut {NoStop}%
\bibitem [{\citenamefont {Verba}\ \emph {et~al.}(2019)\citenamefont {Verba},
  \citenamefont {Tiberkevich},\ and\ \citenamefont {Slavin}}]{Verba2019}%
  \BibitemOpen
  \bibfield  {author} {\bibinfo {author} {\bibfnamefont {R.}~\bibnamefont
  {Verba}}, \bibinfo {author} {\bibfnamefont {V.}~\bibnamefont {Tiberkevich}},\
  and\ \bibinfo {author} {\bibfnamefont {A.}~\bibnamefont {Slavin}},\ }\href
  {https://doi.org/10.1103/PhysRevB.99.174431} {\bibfield  {journal} {\bibinfo
  {journal} {Physical Review B}\ }\textbf {\bibinfo {volume} {99}},\ \bibinfo
  {pages} {174431} (\bibinfo {year} {2019})}\BibitemShut {NoStop}%
\end{thebibliography}%
\clearpage
\appendix

\end{document}


\preprint{AIP/123-QED}
\setlength{\parindent}{0pt}

\title[Supplement Material: Parametric Excitation and Instabilities of Spin Waves driven by Surface Acoustic Waves]{Supplement Material: Parametric Excitation and Instabilities of Spin Waves driven by Surface Acoustic Waves}
\author{Moritz Geilen}
\affiliation{Fachbereich Physik and Landesforschungszentrum OPTIMAS, Technische Universit\"at Kaiserslautern, Germany}
\email{mgeilen@physik.uni-kl.de}
\author{Roman Verba}
\affiliation{Institute of Magnetism, Kyiv 03142, Ukraine}
\author{Alexandra Nicoloiu}
\affiliation{National Institute for Research and Development in Microtechnologies, Bucharest R-07719, Romania}
\author{Daniele Narducci}
\affiliation{imec, Leuven B-3001, Belgium}
\affiliation{KU Leuven, Departement Materiaalkunde, 3001 Leuven, Belgium}
\author{Adrian Dinescu}
\affiliation{National Institute for Research and Development in Microtechnologies, Bucharest R-07719, Romania}
\author{Milan Ender}
\affiliation{Fachbereich Physik and Landesforschungszentrum OPTIMAS, Technische Universit\"at Kaiserslautern, Germany}
\author{Morteza Mohseni}
\affiliation{Fachbereich Physik and Landesforschungszentrum OPTIMAS, Technische Universit\"at Kaiserslautern, Germany}
\author{Florin Ciubotaru}
\affiliation{imec, Leuven B-3001, Belgium}
\author{Mathias Weiler}
\affiliation{Fachbereich Physik and Landesforschungszentrum OPTIMAS, Technische Universit\"at Kaiserslautern, Germany}
\author{Alexandru M\"uller}
\affiliation{National Institute for Research and Development in Microtechnologies, Bucharest R-07719, Romania}
\author{Burkard Hillebrands}
\affiliation{Fachbereich Physik and Landesforschungszentrum OPTIMAS, Technische Universit\"at Kaiserslautern, Germany}
\author{Christoph Adelmann}
\affiliation{imec, Leuven B-3001, Belgium}
\author{Philipp Pirro}
\affiliation{Fachbereich Physik and Landesforschungszentrum OPTIMAS, Technische Universit\"at Kaiserslautern, Germany}

\date{\today}

\maketitle
\section{Time-resolved Measurement}
\begin{figure}[hbt]
	\includegraphics[width=8cm]{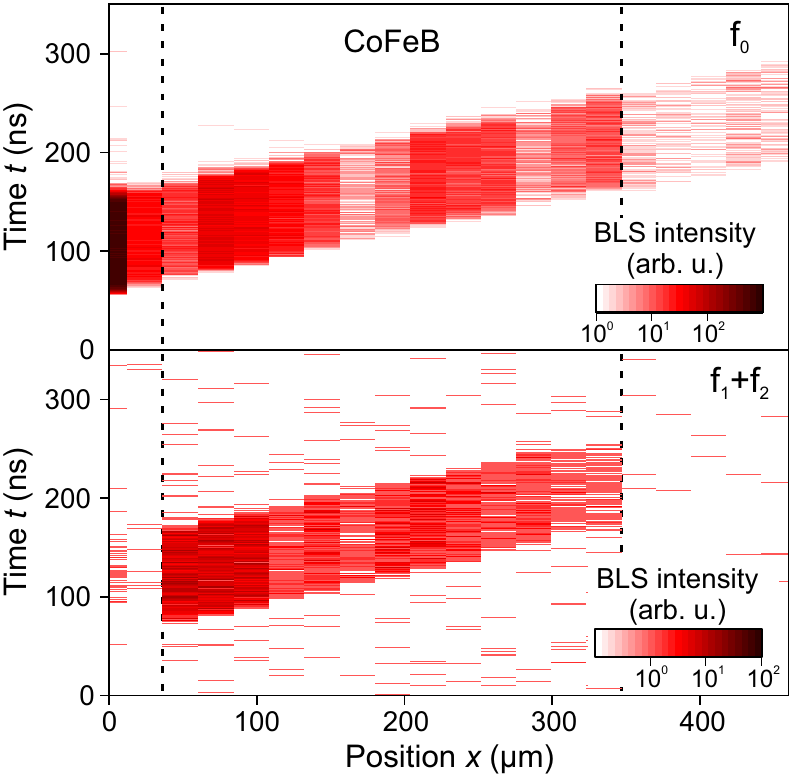}
	\caption{Time-resolved $\upmu$BLS measurements along the propagation direction of the SAWs at a applied field of $\unit[2]{mT}$. The upper panel shows the intensity for $f_0$, while the lower panel shows the integrated signal for the modes $f_1$ and $f_2$.}
	\label{fig:fig4}
\end{figure}

In order to prove that the instability processes are generated by the SAWs and not caused by electromagnetic leakage, time-resolved $\upmu$BLS measurements have been performed. SAW pulses with a pulse duration of $\tau=\unit[100]{ns}$ and an RF-power of $\unit[+30]{dBm}$ were excited and tracked along their propagation. In the upper panel of Fig.~\ref{fig:fig4}, the extracted BLS intensity around $f_0$ is displayed, while in the lower panel, the integrated signal of $f_1$ and $f_2$ is shown. Due to some residual BLS signal from the SAWs, the first is visible throughout the whole distance, while the signal produced by the instability process only appears on the magnetic material. Despite the exponential decay of the SAWs, the amplitude of the magneto-elastic field is large enough to overcome the instability threshold until the end of the magnetic rectangle. The temporal coincidence of the two signals proves that the instability processes are driven by the magneto-elastic field and not by electromagnetic leakage, which would be much faster and would appear instantly over the whole distance.

\section{Micromagnetic Simulations}
In the following, the details for the micromagnetic simulations will be presented. The simulated magnetic pads have a size of $\unit[70]{\mu m}\times\unit[70]{\mu m}\times\unit[18]{nm}$ and were divided into 1024x1024x1 cells. In addition, periodic boundary conditions are assumed. The width of the SAW beam is $\unit[50]{\mu m}$. The following parameters have been used: saturation magnetization $M_\mathrm{S}=\unit[950]{kA/m}$, exchange stiffness $A_\mathrm{ex}=\unit[15]{pJ/m}$, Gilbert damping parameter $\alpha=\unit[0.0043]{}$ and the magneto-elastic coupling constants $B_1=B_2=\unit[-8.8]{MJ/m^3}$ \cite{Vanderveken2021}. Further an uniaxial anisotropy along the x-axis has been assumed with $K_\mathrm{u1}=\unit[1600]{J/m^3}$ \cite{geilen2021fully}. The angle between the x-axis and the external magnetic field is $\varphi=\unit[45]{{}^\circ}$.
The surface acoustic wave has been implemented as two plain waves of the strain components $S_{xx}$ and $S_{xz}$ which are in phase quadrature:
\begin{equation}
    \begin{split}
        S_{xx} & = A \sin(2 \pi (x/\lambda_\mathrm{R}-f t))\\
        S_{xz} & = A \cos(2 \pi (x/\lambda_\mathrm{R}-f t))\\
        S_{yy} & = S_{xy} = S_{zz} = S_{yz} = 0.
    \end{split}
\end{equation}
The wave vector is set by the wavelength $k_\mathrm{R}=\frac{2 \pi}{\lambda_\mathrm{R}}$ with $\lambda_\mathrm{R}=\unit[682]{nm}$. Further, a thermal field corresponding to the temperature $T=\unit[300]{K}$ is used in order to account for the thermal occupation of the magnon modes.\\
\begin{figure*}[bht!]
	\includegraphics[width=16cm]{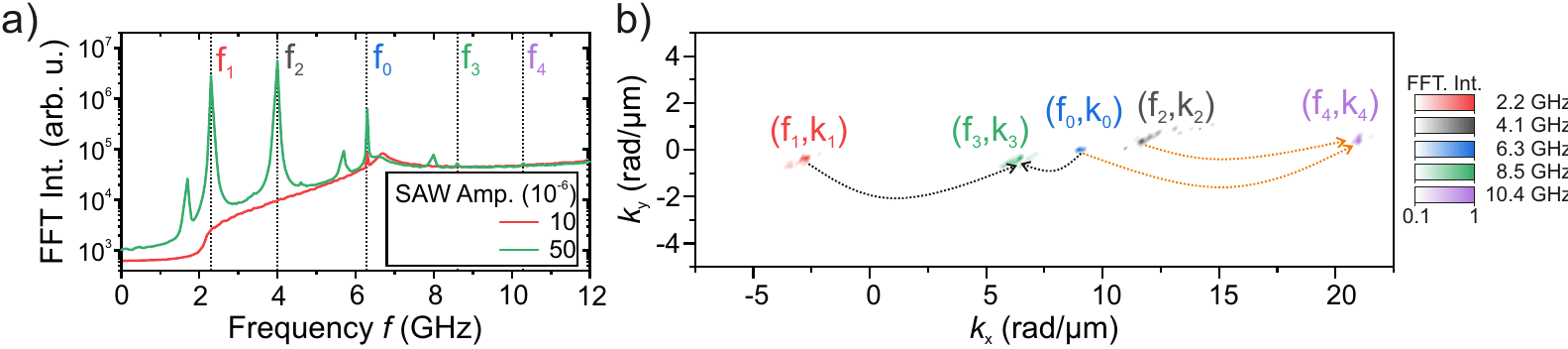}
	\caption{a) The FFT intensity as a function of the frequency extracted from the micromagnetic simulation for $\upmu_0 H_\mathrm{ext}=\unit[2]{mT}$. The linear case is shown in red and the nonlinear in green. b) Extracted FFT intensities for all secondary modes involved in the confluence process leading to $f_3$ and $f_4$ }
	\label{fig:sup1}
\end{figure*}

Figure~\ref{fig:sup1}~a shows the spectra obtained from the micromagnetic simulation for an external field of $\upmu_0 H_\mathrm{ext}=\unit[2]{mT}$. For a strain amplitude of $A=\unit[10\times 10^{-6}]{}$ the spin wave spectrum is dominated by the thermal spin waves. Only a minor linear excitation can be seen at $f_0$. For the nonlinear case with a strain amplitude of $A=\unit[50\times 10^{-6}]{}$ secondary modes appear. The intensity of the pumped modes $f_1$ and $f_2$ exceeds that of the initial mode $f_0$ since the latter is a non-resonant, forced excitation.

The FFT intensity extracted for the frequencies of the initial and secondary modes is shown in Fig.~\ref{fig:sup1} b. The confluence process $f_3 = f_0+f_1$ and $f_4 = f_0+f_2$ fulfill both energy and momentum conservation. 
\noindent

\section{Analytical analysis of SAW-driven spin-wave instabilities}
In the following the analytical modelling of the SAW-driven spin-wave instabilities are discussed. The goal is to determine the threshold values $a_\mathrm{th}$ for the different instability processes and to determine the minimum threshold value $a_\mathrm{th,min}$.  This value indicates which scattering process starts first and which secondary spin-wave modes are populated as a result.\\
The spin-wave dispersion in an in-plane magnetized film with uniaxial anisotropy in $x$-direction is given by:
\begin{equation}
  \begin{split}
    \omega_k^2 & = \Omega_\mathrm{IP} \Omega_\mathrm{zz} = \left(\omega_\mathrm{H} + \omega_\mathrm{M} \lambda^2 k^2 + \omega_\mathrm{M} F_{y'y'} \right) \\
    \times &\left(\omega_\mathrm{H} + \omega_\mathrm{M} \lambda^2 k^2 + \omega_\mathrm{M} (1-f(kh)) \right) \,,
    \end{split}
\end{equation}
with 
  \begin{equation}
   F_{y'y'} = \frac{k_\bot^2}{k^2}f(kh) - \frac{B_\mathrm{a}}{\mu_0 M_\mathrm{S}} \sin^2 \phi_\mathrm{M} \,
  \end{equation}
and $f(x) = 1-(1-e^{-|x|})/|x|$, where $k_\bot = k_y \cos \phi_\mathrm{M} - k_x \sin\phi_\mathrm{M}$ is the spin-wave wave vector component perpendicular to the static magnetization. The anisotropy field is given by $B_\mathrm{a} = 2K_\mathrm{u}/M_\mathrm{S}$. The parameters used are consistent with those used in the micromagnetic simulations. For the sake of brevity we will use the following notation for the spin-amplitude with the wavevector $\mathbf{k}_1$: $c_{\mathbf{k}_1} \equiv c_\mathbf{1}$.\\

\begin{figure}[bht!]
	\includegraphics[width=8cm]{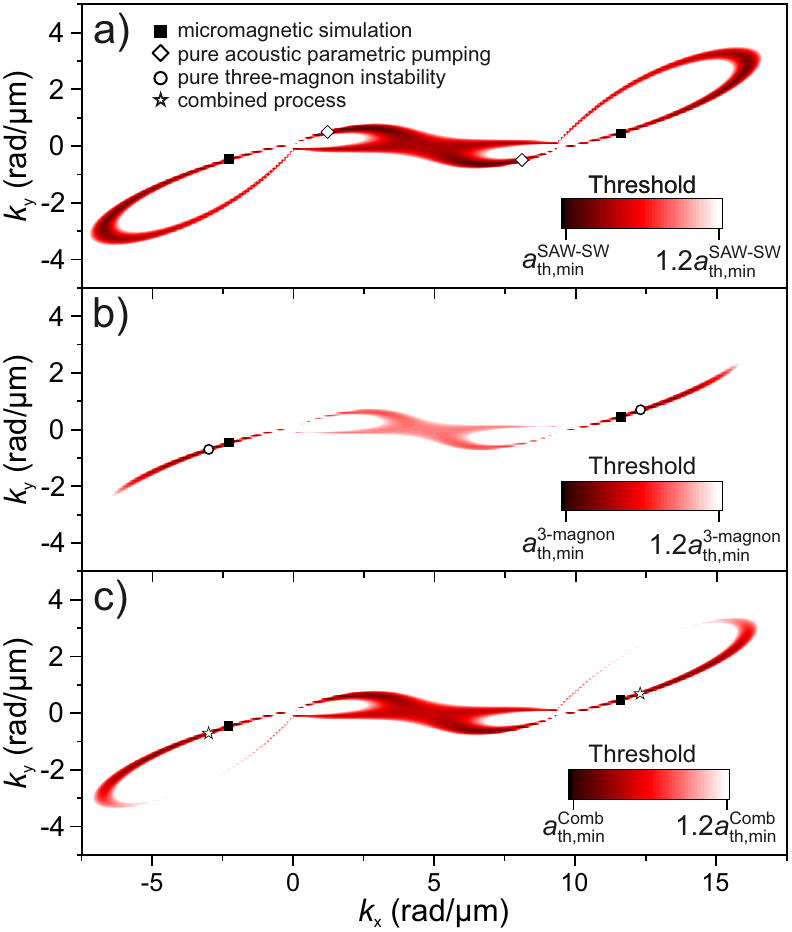}
	\caption{Calculated threshold values for a) the direct SAW-SW process of the acoustic pumping process, b) the first-order spin wave instability of non-resonantly excited SWs, c) the combined process of direct SAW-SW and three-magnon splitting. The spin-wave modes with minimal threshold values for the direct SAW-SW process $a_\mathrm{th,min}^\mathrm{SAW-SW}$ (diamond), the first order spin-wave instability of non-resonantly excited SW (Circle) and the combined process $a_\mathrm{th,min}^\mathrm{Comb}$ (Star) are indicated. The positions of those modes extracted from the micromagnetic simulations are shown by the black squares.}
	\label{fig:sup4}
\end{figure}

A theory of parametric interaction of spin waves and surface acoustic waves including acoustic pumping was developed in \cite{Lisenkov2019}. The dynamic equations for the spin-wave amplitudes $c_\mathbf{k}$ read:
\begin{equation}
   \frac{dc_\mathbf{1}}{dt} + i \omega_\mathbf{1}  c_\mathbf{1} + \Gamma_\mathbf{1} c_\mathbf{1} = i \mathcal{V}_{\mathbf{0},\mathbf{1} \mathbf{2}} c_{\mathbf{2}}^* a_{\mathbf{SAW}} e^{-i\omega_\mathrm{SAW} t}  \,,
\end{equation}
where $a_{\mathbf{SAW}}$ is the complex SAW amplitude and $\Gamma_\mathbf{k}$ is spin-wave damping rate. The spin-wave (SW) and SAW wave vectors satisfy the relation $\mathbf{k_0} = \mathbf{k_1} + \mathbf{k_2}$ and the parametric coupling efficiency is given by:
  \begin{equation}\label{e:V_SAW}
   \mathcal{V}_{\mathbf{0}, \mathbf{1} \mathbf{2}} = \frac{2\gamma B_1}{M_S}\bar S_{\mathbf{0}, xx} \left(m_{\mathbf{1},x}^* m_{\mathbf{2},x}^* - \mathbf{m}_{\mathbf{1}}^* \cdot \mathbf{m}_{\mathbf{2}}^* \cos^2(\phi_M) \right) \ . 
  \end{equation}
Here, $\bar{S}_{\mathbf{0}, xx}$ is the thickness average of the SAW strain $S_{\mathbf{0}, xx}$ (the physical strain is related to the SAW amplitude by $S_{ij}(\mathbf{r}, t) = a_\mathbf{0} S_{\mathbf{0}, ij}(z) \exp({ik_0 x - i\omega_\mathrm{SAW} t}) +\mathrm{c.c.} $). The normalized spin-wave profile is given by $\mathbf{m}_\mathbf{k} = [-m_\mathrm{IP} \sin (\phi_M), m_\mathrm{IP} \cos(\phi_\mathrm{M}), m_z ]$ with $m_\mathrm{IP} = \sqrt{\Omega_\mathrm{zz}/2\omega_k}$ and $m_z = i\sqrt{\Omega_\mathrm{IP}/2\omega_k}$.\\
In the following we will disregard the effect of the off-diagonal SAW strain components $S_{xz} = S_{zx}$, since in the considered case this strain component, averaged over the ferromagnetic layer, is more than 10 times smaller than the normal strain component $S_{xx}$. In addition a careful analysis has shown that the off-diagonal contributions do not lead to any qualitatively new features. The other assumption made is that the spin-wave profile across the ferromagnet thickness is assumed to be uniform. 
It is noteworthy that in Eq.~\ref{e:V_SAW} two contributions relate to the parametric coupling process. The first term which is defined by the square of dynamic magnetization component, appears only in the case of anisotropy-type pumping, where the pumping field itself depends on the magnetization as it is the case with  e.g., for a magneto-elastic \cite{Lisenkov2019} or a magneto-electric \cite{Verba2014} drive. The second term is proportional to the spin-wave ellipticity, like in standard microwave parallel pumping. In the general case of a non-resonant process the threshold of SAW-SW three-wave splitting is equal to: \cite{Verba2021}
\begin{equation}\label{e:th-dir}
    a_\mathrm{th} = \frac{1}{|\mathcal{V}_{\mathbf{0},\mathbf{1}\mathbf{2}}|} \sqrt{\Gamma_1 \Gamma_2 \left(1+ \frac{\delta\omega^2}{(\Gamma_1 + \Gamma_2)^2} \right)} \,,
\end{equation}
where $\delta\omega = \omega_\mathrm{SAW} - (\omega_1 + \omega_2)$ is the detuning from the three-wave resonance splitting condition. The values for the threshold of this acoustic pumping process are shown in Fig.~\ref{fig:sup4}. Small thresholds, naturally, correspond to a resonance condition ($\omega_\mathrm{SAW} = \omega_1 + \omega_2$), at which, the dependence of the threshold value is quite pronounced. Using this approach the minimum threshold for the acoustic pumping is calculated to $a_\mathrm{th,min}^{\mathrm{SAW-SW}}=\unit[3.7 \times 10^{-5}]{}$ (respective amplitude of the SAW strain is $|S_{xx,\mathrm{th}}| = 2 a_\mathrm{th,min}^{\mathrm{SAW-SW}}=\unit[7.4 \times 10^{-5}]{}$), which corresponds to the splitting into spin-wave modes with $\mathbf{k_1} = \unit[(8.1, -0.5)]{rad/\mu m}$ and $\mathbf{k_2} = \unit[(1.2, 0.5)]{rad/\mu m}$ (see diamonds in Fig.~\ref{fig:sup4}).\\

Now, the above presented approach is adapted to the three-magnon splitting of non-resonantly excited SWs at $\{f_\mathrm{SAW}, k_\mathrm{SAW} \}$, which is the second possible mechanism that leads to the population of the spin-wave modes $f_1$ and $f_2$. The expression for the splitting threshold is given by Eq.~\eqref{e:th-dir}, where the threshold SAW amplitude $a_\mathrm{th}$ is now replaced by the spin-wave amplitude $c_0$, and the coefficient $\mathcal{V}_{\mathbf{0},\mathbf{1}\mathbf{2}}$ is replaced by the three-magnon scattering coefficient $V_{\mathbf{0},\mathbf{1}\mathbf{2}}$. The latter is straightforwardly calculated using the Hamiltonian formalism \cite{Krivosik2010, Verba2019}. An explicit expression is rather cumbersome and is not presented here. The relation of the spin-wave amplitude $c_0$ with the SAW amplitude due to the magneto-elastic coupling is given by:
\begin{equation}
   c_\mathbf{0} = \frac{i\gamma b_\mathbf{0} a_\mathbf{0}}{i(\omega_\mathbf{0} - \omega_\mathrm{SAW}) + \Gamma_\mathbf{0}}
\end{equation}
with the linear coupling efficiency
\begin{equation}
   b_\mathbf{k} = \frac{2B_1}{M_\mathrm{S}} m_{\mathbf{k},x}^* \bar S_{\mathbf{k},xx} \cos\phi_\mathrm{M} \ .
\end{equation}
The calculation of the minimum threshold for this three-magnon process $a_\mathrm{th,min}^\mathrm{3magnon}$  reveals that secondary spin waves with $\mathbf{k}_1 = \unit[(12.3, 0.1)]{rad/\mu m}$ and $\mathbf{k}_2 = \unit[(-3.0, -0.1)]{rad/\mu m}$ are excited. However, the magnitude of $a_\mathrm{th,min}^\mathrm{3magnon}=\unit[6.05 \times 10^{-5}]{}$ is significantly (64\%) larger compared to the minimal threshold of the acoustic pumping.\\

As explained in the main text both scattering mechanism take place even below the threshold of the instability. Therefore, both mechanisms have to be taken into account when calculating the minimal threshold of the real system. The minimum threshold of the combined process ($a_\mathrm{th,min}^\mathrm{Comb}=\unit[3.15 \times 10^{-5}]{}$) is reached at $\mathbf{k}_1 = \unit[(-3, -0.7)]{rad/\mu m}$, $\mathbf{k}_2 = \unit[(12.3, 0.7)]{rad/\mu m}$ (see~Fig.~\ref{fig:sup4}, circle). This threshold is due to the additional energy transferred from the three-magnon process lower than for the pure acoustic pumping. The calculated threshold for the wave vectors obtained from micromagnetic simulations is $a_\mathrm{th,Simulation}^\mathrm{Comb}=\unit[3.34 \times 10^{-5}]{}$. The position of those modes is indicated by the black squares in Fig.~\ref{fig:sup4}. Calculated minimal threshold SAW strain amplitude $S_{xx,\mathrm{th}}^\mathrm{Comb} = 2 a_\mathrm{th,min}^\mathrm{Comb}=\unit[6.3 \times 10^{-5}]{}$ is somewhat higher than one observed in micromagnetic simulations ($5\times10^{-5}$) because of impact of $S_{xz}$ strain, accounted in simulations. \\ 

Finally, the processes leading to the population of the modes $f_3$ and $f_4$ will be discussed. These processes can be either three-wave confluence processes (spin-wave confluence or SAW-spin wave confluence) or four-wave splitting processes. An important observation is that the states at $f_3$ and $f_4$ are not linear eigenmodes of the spin-wave system. This implies that they cannot be a result of a spontaneous splitting process. Although non-resonant splitting is possible, both secondary states need to be non-resonant. The situation when one secondary state is resonant ($f_{1,2}$ in our case), while the other one is non-resonant, is not allowed \cite{Verba2021}. Therefore, the four-wave process must be stimulated by spin waves at $f_{1,2}$.\\

In order to compare the magnitude of three-wave and four-wave terms a Hamiltonian approach is used. In the dynamical equations these terms appear as
  \begin{equation}
    \frac{dc_\mathbf{3}}{dt} \sim \mathcal{V}_{\mathbf{3}, \mathbf{0}\mathbf{1}}^* a_\mathbf{0} c_\mathbf{1} + V_{\mathbf{3}, \mathbf{0}\mathbf{1}}^* c_\mathbf{0} c_\mathbf{1} + W_{\mathbf{0} \mathbf{0}, \mathbf{2}\mathbf{3}} c_\mathbf{0}^2 c_\mathbf{2} \,,
  \end{equation}
where $W_{ij,kl}$ is the four-magnon coefficient and $\mathcal{V}_{\mathbf{3}, \mathbf{0}\mathbf{1}}$ is the three-wave coefficient of the SAW+SW$\to$SW process. In fact, one needs to compare the values of $\mathcal{V}_{\mathbf{3}, \mathbf{0}\mathbf{1}} (dc_0/da_0)$, $V_{\mathbf{3}, \mathbf{0}\mathbf{1}}^*$ and $W_{\mathbf{0} \mathbf{0}, \mathbf{2}\mathbf{3}} c_\mathbf{0}$. The calculations made using the Hamiltonian formalism gives $V\sim\unit[2\pi\times 18]{GHz}$ and $W \sim\unit[2\pi\times(25-75)]{GHz}$ for the different processes. Using a SAW-SW perturbation approach we get $\mathcal{V} (dc/da) \sim\unit[2\pi\times 17]{GHz}$. The values of the directly excited mode amplitude is known from the threshold of the first three-wave splitting process, $c_\mathbf{0} \approx 0.0045$. Thus, in all the studied range (one order of magnitude above the threshold value) the four-magnon process is about two orders of magnitude weaker than the three-magnon processes. In addition, it can be concluded that the SAW-SW confluence process is of the same order as the three-magnon confluence process and both these processes contributes to the appearance of signals at $f_{3,4}$ (please note, that confluence processes are non-threshold).
\bibliographystyle{apsrev4-2}
\bibliography{Supplement}